\documentclass[lettersize,journal]{IEEEtran}
\usepackage{amsmath,amsfonts}
\usepackage{algorithmic}
\usepackage[linesnumbered, ruled]{algorithm2e}
\usepackage{array}
\usepackage[caption=false,font=normalsize,labelfont=sf,textfont=sf]{subfig}
\usepackage[font=normalsize,labelfont=bf]{caption}
\usepackage{textcomp}
\usepackage{stfloats}
\usepackage{url}
\usepackage{verbatim}
\usepackage{graphicx}
\usepackage{cite}
\usepackage{xcolor}
\usepackage{textcomp}

\begin{document}

\title{Improving Instruction Fetch Efficiency via High-Level Program Map Traversal} %\textcolor{green}{Sequencing}}

\author{Shyam Murthy, Gurindar S Sohi\\
University of Wisconsin-Madison\\
\{shyamm, sohi\}@cs.wisc.edu}

\maketitle

\begin{abstract}

%\textcolor{red}{This paper proposes a high-level program sequencing scheme that can sequence to resolve upcoming high-level control flow. Further, this paper studies one application of this technique to faciliate efficiency in the instruction fetching process.} 

Efficiency in instruction fetching is critical to performance,
and this requires the primary structures---L1 instruction caches (L1i),
branch target buffers (BTB) and instruction TLBs (iTLB)---to have the requisite information
when needed. This paper proposes 
\textit{instruction presending}, which traverses a high-level program map to identify
%a high-level program \textcolor{green}{sequencing} mechanism and a 
%coupled technique for block movement, 
and move instruction cache blocks, BTB entries, and iTLB entries from the secondary to the primary structures in a "just in time" fashion.
%so that they are available when needed.

Empirical results are presented to demonstrate the efficacy of the proposed presending scheme.
%high-level \textcolor{green}{sequencing} mechanism and block movement.  
%Presending is especially effective for benchmarks with a high base MPKI, where the movement of instruction blocks (and BTB/iTLB entries)  from secondary to primary structures is frequent.
Presending reduces the number of cycles where the instruction fetch is waiting by an order of magnitude as compared to state-of-the-art instruction prefetching schemes while operating with small-sized primary BTBs.
It is especially effective
for benchmarks with a high base MPKI, where movement 
%of instruction blocks (and BTB/iTLB entries) 
from secondary to primary structures is frequent.
This improvement in fetch efficiency results in performance improvements in cases where this efficiency is important. 

%. \textcolor{red}{Presending can do so with small sized primary BTBs (why we replicate some information already present in processor front-ends).}

%is critical to a processor's performance and efficiency,
%a variety of \textit{instruction prefetching} schemes have been proposed to \textit{prefetch}
%cache blocks holding instructions from lower levels of a memory hierarchy into the L1 instruction cache.

\end{abstract}

\section{Introduction}

Processor front-ends are designed around the idea of a processor fetching instructions from memory hierarchies, with multiple levels of caches, and additional structures such as \textit{instruction TLBs (iTLBs)} and \textit{branch target buffers (BTBs)}, used to support and speed up the fetching process. Many modern applications have active code footprints that are too large to achieve high hit ratios even with larger (primary) L1i caches, yet fit almost entirely in cache sizes that are common in on-chip (secondary) L3 caches
\cite{ayers2019asmdb,ayers2018memory,sriraman2019softsku,vavouliotis2021morrigan,song2022thermometer,brunner2024weeding,wang2023acic,kanev2015profiling,ailamaki1999dbmss,keeton1998performance,cao1999detailed}.
A similar situation occurs for relatively small-sized primary iTLBs and BTBs.
Overall, demand missing in primary structures (L1i cache, L1 iTLB, L1 BTB), and movement from secondary structures, is frequent. The question that arises then is: how to proactively carry out the (frequent) movement from one part of a processing chip (the secondary structures) to another (the primary structures)?

%The problem that we face is how to get the instruction cache blocks (and BTB and iTLB entries)
%from the secondary structures where they reside to the primary structures where they are needed by the processor,
%i.e., move them from one part of a processing chip to another.

%This results in the processor frequently missing in the L1 instruction cache
%and demand fetching, or prefetching, to bring code blocks from elsewhere on the chip (the L2/L3) into the L1.

%The expectation is that instruction fetch requests would be serviced by the L1 instruction cache,
%since a delay in fetching instructions results in a significant loss of performance.

Prefetching is used to carry out the movement from the secondary to the primary structures, and a plethora of \textit{instruction prefetching} schemes have been proposed
\cite{zhang2002execution,spracklen2005effective,srinivasan2001branch,reinman1999fetch,ferdman2008temporal,ferdman2011proactive,kolli2013rdip,kaynak2015confluence,kumar2018blasting,kumar2017boomerang,ansari2020divide,ishii2021re,ros2021cost, ayers2019asmdb,annavaram2003call,soundararajan2021pdede}.
Several prefetching schemes are closely coupled to the processor front-end and rely on the processor's low-level (branch-by-branch) sequencing through the program to identify the (precise) upcoming dynamic instruction stream. This low-level sequencing is impacted by the effectiveness of
\textit{branch direction predictors} and BTBs.
Other prefetching schemes learn about blocks likely to miss and tie them to instruction stream events. Additionally, software-based prefetching strategies insert prefetch instructions into the program, typically informed by profiling or offline analysis \cite{song2022thermometer,ayers2019asmdb,khan2021twig}.

%\cite{hsu1992prefetching,smith1982cache,park1997improved,zhang2002execution,spracklen2005effective,srinivasan2001branch,ferdman2008temporal,ferdman2011proactive,kolli2013rdip,kaynak2015confluence,1410067,ansari2020divide,ros2021cost}.

%For example, a plethora of \textit{instruction prefetching} schemes, which are typically intimately tied to a processor's fetch process,
%have been proposed to prefetch cache blocks into the L1i cache 

%Getting information into the L1 cache, BTB, iTLB, is via a process that is intimately tied to the fetch process of  a processor
%and the (mostly) precise sequence of program instructions executed by the processor.

%This begs the question: are instruction prefetching concepts, introduced when it took multiple chips to build a processor, 
%and are closely tied to the advanced discovery of the (mostly) precise sequence of instructions to be executed,
%the best way to move information between components of a memory hierarchy on the same chip?  For example, from a large L3 cache
%to a small L1 cache, or from a large BTB to a small BTB?

%\textit{instruction presending}, where the cache blocks holding (static) instructions, and associated BTB and iTLB information,
%are pulled from the larger, slower, secondary structures on chip, and sent to the (potentially very) smaller, faster, primary structures (L1i cache, BTB, iTLB) close to
%the processor, in a "just in time" fashion.

This paper presents an alternate mechanism to identify the subset of the static code that the processor is likely to be processing imminently, and proactively move the code blocks, as well as the BTB and iTLB entries corresponding to the instructions in those blocks,
%blocks of instructions (and the BTB/iTLB entries for instructions in those blocks) 
%likely to be imminently needed by the processor, and proactively move them 
for that subset from secondary to primary structures.
The mechanism, \textit{instruction presending}, works with a \textit{high-level program map}, derived from the program call graph.
%which is created from the static program.
%The high-level program map 
This map maintains information about the instruction cache blocks, BTB, and iTLB entries for \textit{fragments} (static code subsets) of a program, and pathways from one fragment to the next in a dynamic execution.
%with each fragment containing information about 
%instruction cache blocks, BTB, and iTLB entries needed to process the instructions in that fragment, and identifiers of successor fragment(s).  
The presending mechanism operates to traverse the high-level program map, identify the instruction cache blocks, BTB and iTLB entries that the processor is going to need imminently to execute the program, and move them from secondary to primary structures in a "just in time" fashion.  Presending does not try to create the precise dynamic instruction stream and thus does not need information from branch predictors and BTBs for its operation.
Presending operates mostly decoupled from the processor, using select information sent from the processor to keep the traversal through the high-level program map on track, and sufficiently ahead of the processor's execution.  With presending, the cycles where the processor front end is waiting for instructions is significantly lower (e.g., order of magnitude), resulting in higher performance, than even the best prefetching schemes, with smaller-sized primary BTBs.

\section{Motivation}
\label{sec:Motivation}
\subsection{Rate of Secondary Structure Involvement}
\label{sec:Rate-Move}

%\begin{figure*}[ht]
 %   \centering
  %  \begin{minipage}[b]{0.55\textwidth}
   %     \centering
    %  \input{tables/mot_table_11.tex}
     %   \caption{Select Benchmark Characteristics}
      %  \label{fig:sig-callgraph-walk}
    %\end{minipage}
    %\hspace{0.02\textwidth} % Space between figure and table
    %\begin{minipage}[b]{0.4\textwidth}
     %   \centering
      %  \includegraphics[width=\textwidth]{paper_diagrams_2/program_snippet_1.pdf}
      %  \caption{Example Program Snippet}
       % \label{fig:program-snippet}
    %\end{minipage}
%\end{figure*}

\begin{table*}[]
	\centering

\caption{Select Benchmark Characteristics}

\begin{tabular}{lllllllllllll}
\textbf{}    & \multicolumn{2}{c}{\textbf{L1i}} & \multicolumn{4}{c}{\textbf{RPKI}}           & \multicolumn{4}{c}{\textbf{CWKI}} & \textbf{Static} & \textbf{\%}       \\\hline\hline
\textbf{}    & \multicolumn{2}{c}{\textbf{MPK}} & \multicolumn{3}{c}{\textbf{FDIP}}          & \textbf{PS}  & \multicolumn{3}{c}{\textbf{FDIP}}           & \textbf{PS}   & \textbf{}       & \textbf{frag}    \\\hline
\textbf{App} & \textbf{I}      & \textbf{A}     & \textbf{512} & \textbf{8K}  & \textbf{$\infty$} & \textbf{}    & \textbf{512}  & \textbf{8k}  & \textbf{$\infty$} & \textbf{}     & \textbf{95th}   & \textbf{(2, $>$2)} \\\hline
\textbf{}    & \textbf{(1)}    & \textbf{(2)}   & \textbf{(3)} & \textbf{(4)} & \textbf{(5)} & \textbf{(6)} & \textbf{(7)}  & \textbf{(8)} & \textbf{(9)} & \textbf{(10)} & \textbf{(11)}   & \textbf{(12)}    \\\hline
s1           & 10              & 97             & 10.1         & 6.0          & 5.8          & 0.7          & 322           & 166          & 135          & 43            & 944             & 9.2,2            \\
s2           & 15              & 121            & 15.0         & 4.6          & 4.6          & 0.1          & 314           & 78           & 77           & 1             & 250             & 2.4,0.1          \\\hline
s10          & 24              & 247            & 20.8         & 7.4          & 5.9          & 1.3          & 544           & 172          & 100          & 15            & 2638            & 7.6,1.8          \\
s13          & 27              & 272            & 22.3         & 7.4          & 5.8          & 1.3          & 602           & 190          & 112          & 17            & 3119            & 7.2,2            \\\hline
s16          & 36              & 438            & 33.0         & 4.9          & 4.6          & 0.2          & 861           & 111          & 93           & 2             & 3132            & 3.9,0.3          \\
s19          & 44              & 438            & 28.3         & 4.4          & 4.4          & 0.1          & 923           & 112          & 112          & 21            & 606             & 4.9,0.8          \\\hline
s23          & 48              & 444            & 44.0         & 6.6          & 6.1          & 0.2          & 1197          & 154          & 126          & 4             & 3151            & 4.4,0.5          \\
s26          & 54              & 475            & 47.7         & 6.7          & 6.1          & 0.2          & 1324          & 176          & 141          & 3             & 3096            & 3.6,0.3          \\\hline
s32          & 63              & 508            & 53.2         & 5.7          & 5.7          & 0.1          & 1538          & 148          & 150          & 1             & 2129            & 1.8,0            \\
s38          & 79              & 609            & 59.7         & 3.4          & 3.4          & 0.0          & 1852          & 129          & 129          & 1             & 631             & 2.2,0           
\end{tabular}

\label{fig:sig-callgraph-walk}
\centering
\end{table*}

%When an instruction fetch misses a primary (L1i) cache, the secondary (L2/L3) cache gets involved.
%A canonical metric for L1i misses is \textit{misses per kilo instructions (MPKI)}.
%However, with multi-issue processors, multiple instructions are fetched with a single L1 fetch request,
%thus \textit{misses per kilo fetch requests (MPKF)} is more representative of the fraction of
%L1 accesses that also involve the L2.
%However, given that a single L1i access fetches multiple instructions, another illustrative measure of the rate of L2 cache involvement is the \textit{misses per kilo cache accesses (MPKA)}.
%Block references are references to different blocks; consecutive accesses to the same block are combined
%into a single block reference.

When an instruction fetch misses in the primary instruction cache (L1i), the secondary cache (L2/L3) must be accessed. The canonical metric for instruction cache misses is \textit{misses per kilo instructions (MPKI)}. However, in modern processors MPKI underestimates the relative L1i miss rate due to two factors: (i) instruction buffers hold the most recently accessed cache line, and instruction fetches served from the buffer don't access the L1i,
(ii) wide-issue processors fetch multiple instructions in a single access.
These factors result in a smaller number of L1i accesses and thus a much higher L1i miss rate for the same number of L1i misses.
Consequently, an additional metric—\textit{misses per kilo accesses (MPKA)}, which quantifies the fraction of L1i accesses that miss and involve a secondary cache access, provides a more direct measure of how frequently the secondary cache is involved when the L1i is accessed. For instance, an MPKA of 500 indicates that roughly one out of every two L1i accesses results in an L2 access.

Table \ref{fig:sig-callgraph-walk} presents some empirical characteristics of a subset of server benchmarks,
with two benchmarks chosen from groups that have a base L1i MPKI of 10-15, 15-30, 30-45, 45-60, and $>$ 60, respectively.
(More details of the benchmarks and simulation setup are in section \ref{sec:Eval}.)
The first two data columns (1 and 2) of Table \ref{fig:sig-callgraph-walk} present the
MPKI, and MPKA for a 32KB, 8-way associative cache.
Even for benchmarks with a relatively low MPKI (e.g., s1), the MPKA is quite high (about 97 per 1000 L1i accesses), i.e., nearly one in ten L1i accesses ends up accessing the L2 cache. For high-MPKI workloads (e.g., s32), the MPKA exceeds 500, indicating that more than half of all L1i accesses result in an L2 access. The large code footprint also leads to more frequent primary iTLB and BTB misses, and consequently a high rate of involvement of secondary TLBs and BTBs.

%Instruction processing also requires information from iTLBs and BTBs.
%The needed iTLB information is uniquely determined by the static code block containing the instruction.
%Similarly, the BTB information needed is closely related to the corresponding static code block.
%The 3rd and 4th data columns present the MPKA for iTLBs with 32 and 128 entries, respectively,
%and the 5th and 6th data columns the MPKA for BTBs with 512 and 2K entries, respectively;
%with 4-way set associativity in each case.

%\textcolor{red}{The above para is not necessary because we don't present MPKI/MPKA for an iTLB and BTB. We need supporting text for cycles waiting for an instruction with different BTB sizes and for Send.  }

\subsection{Prefetching for Block Movement}
\label{sec:Prefetch-Move}

Instruction prefetching techniques 
%some of which we will discuss in section \ref{sec:Rel-work},
try to get sufficiently ahead of the fetch process and timely prefetch needed blocks into the L1i cache. Similarly, entries can be prefetched into BTBs; prefetching for iTLBs is not common because of
the risk of pollution in extremely small-sized structures.
The overall goal is to 
%\textit{improve the efficiency} of the fetch process, i.e.,
\textit{reduce the number of cycles that the fetch process is waiting}.

Many prefetching techniques, of which \textit{Fetch Directed Instruction Prefetching (FDIP)} is a practical example,
%especially BTB/Fetch directed prefetchers \textcolor{red}{\textit{FDIP}}) 
rely on the identification of the precise instruction fetch stream.
To proceed past branches, they need a correct branch direction (via a branch direction predictor) and a correct branch target (via a BTB).
This has led to the need for very large BTBs
(e.g., the IBM Z15 \cite{adiga2020ibm} has a 16K entry primary BTB and a 128K entry secondary BTB) or innovations in BTB design 
%to achieve high hit rates without a brute-force increase in size, and 
such as combining instruction/BTB prefetching and BTB design, e.g.,
Confluence \cite{kaynak2015confluence} and Shotgun \cite{kumar2018blasting}.

%Fetch/prefetch techniques employ the best branch direction predictor that they can, along with an adequate BTB.

\begin{figure}[h]
   \centering
  \includegraphics[scale=0.8]{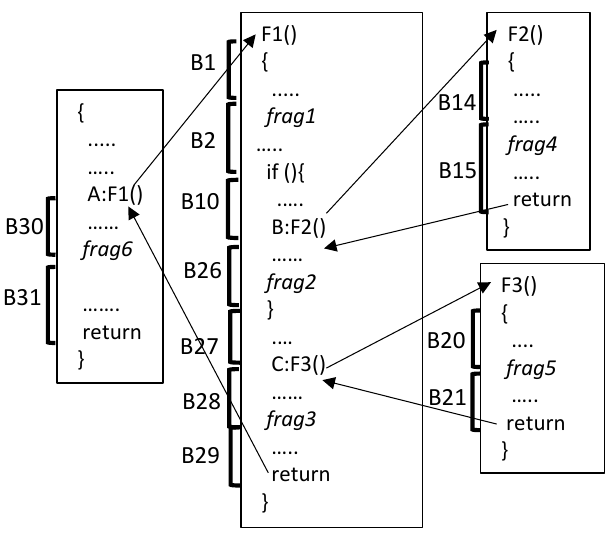}
	\caption {Example Program Snippet}
  \label{fig:program-snippet}
\end{figure}

A \textit{redirect} of the fetch process happens when it is determined that fetch was proceeding down an incorrect
path due to a branch mispredict or incorrect branch target.
A redirect of fetch typically also redirects the associated prefetch process.
The 3rd, 4th, and 5th columns present the \textit{redirects per kilo-instruction (RPKI)} for FDIP for branch target buffers (BTBs) with 512, 8K, and $\infty$ entries, respectively.
The 7th, 8th, and the 9th columns present the corresponding \textit{cycles waiting per kilo-instruction (CWKI)} that the fetch process is waiting because the desired instruction block isn't in the L1i cache.
%\textcolor{red}{The 3rd, 4th, and 5th columns present the processor cycles spent waiting for instructions per kilo-instruction (KI) for branch target buffers (BTBs) with 512, 8K, and infinite entries, respectively. }

Observe a very high CWKI (exceeding 1000) with a 512 BTB due to a high RPKI. The 8K BTB reduces RPKI (and consequently CWKI) significantly in most cases.
With an $\infty$ BTB, FDIP is redirected only on branch direction mispredicts.  The RPKI is in the single digits and the CWKI is about 100. Both RPKI and CWKI will be significantly reduced, as shown in the 6th and 10th columns, respectively, by our proposed presending mechanism (PS) which traverses a high-level program map to identify the needed instruction blocks and moves them to the primary structures.
%informationthrough high-level program sequencing and block movement.

%in waiting cycles, as reflected in the 8th and 4th columns. While an infinite-entry BTB provides only marginal improvements in both redirect frequency and waiting cycles compared to the 8K-entry BTB, there is still room for improvement, as non-trivial waiting cycles, often in the hundreds, persist. The remaining inefficiency with an infinite-entry BTB is primarily due to direction mispredicts.

%The 7th and 8th data columns present the redirects per kilo instructions (KI) for BTBs with 2K and 8K entries, respectively.

\subsection{High-Level Program Structure and Map}
\label{sec:High-Level-Structure}

Many modern programs are highly structured, with many different functions,
and the call graph of the program represents the flow of execution through the program at a function level. Within each function there are \textit{fragments} of code: 
a \textit{fragment} is a set of blocks of code that start at a target of a call/return instruction,
and end at the next (dynamic) call/return instruction. 

Figure \ref{fig:program-snippet} illustrates an example. Function F1 is called at site A, beginning with code cache blocks B1, B2, and B10, which form fragment \textit{frag1}. Then, function F2 (\textit{frag4}) may be called at site B, depending on an if condition, and has code blocks B14 and B15, while function F3 (\textit{frag5}) is called at site C, and has code blocks B20 and B21. Thus, if the program starts at \textit{frag1}, the next fragment is either \textit{frag4} (if F2 is called) or \textit{frag5} (if F3 is called). When the code in {frag4} completes execution, execution returns to \textit{frag2}, which contains code blocks B26 and B27. Similarly, the next fragments following \textit{frag2} and \textit{frag3} are \textit{frag5} and \textit{frag6}, respectively.

The 11th and 12th columns 
%of Table \ref{fig:sig-callgraph-walk} 
present characteristics of fragments in the selected benchmark programs.
The 11th column has the number of static fragments that account for 95\% of the execution,
and the 12th has a tuple with the percentage of dynamically executed fragments that have two, or more than two possible next fragments.
%\textbf{fragment terminators} (change it to this),
%and the last data column has the average number of cache blocks in a fragment.
For example, for s10, 2638 static fragments account for 95\% of the execution, and
7.6\% and 1.8\% of the dynamic fragments have two and more than two next fragments, respectively,
the remaining 90.6\% having a single next fragment.

The \textit{high-level program map}, or simply \textit{program map}, that we use for instruction presending in this paper is the fragment-level representation of the program's call graph as above.  This map is maintained in a table, with an entry in the table for each fragment. To make an analogy with a traditional road map, the fragments are cities, the information for each fragment corresponds to the roads in the city, and the next fragments correspond to the highways to the next city/cities.
The data of column 11 suggests that this program map could be maintained in tables of a few thousand entries.

\subsection{Instruction Presending for Block Movement}
\label{sec:Presend-Structure}

%With a high rate of block movement 
%(section \ref{sec:Rate-Move}),
%a technique where blocks are 
%proactively sending blocks from secondary to primary structures is worth considering.

The main idea of \textit{Instruction Presending (PS)} is to have
an \textit{Instruction Presending Unit (IPU)} traverse the program map at a fragment level, identify the blocks that the processor needs imminently,
%from the information provided in the map for each fragment, 
and proactively move them to the primary structures.
In traversing the map from one fragment to the next, the IPU only needs to know the successor fragment(s) (which city/cities come next), and does not need to know how execution proceeds through the code blocks in a fragment (which path is taken through the roads of a city).
%, and thus does not need branch predictors or BTBs that a processor needs to determine the precise path of execution through the code blocks of a fragment. 
Data of column 12 suggests that most of the time the IPU would only have a single successor fragment.

The IPU tracks the processor's progress on program execution to see if the IPU is proceeding correctly, i.e, the processor is reaching the same sequence of fragments, and is
\textit{redirected} if it is proceeding incorrectly.
%in its high-level sequencing.
%As we shall see, by selectively proceeding along up to two paths, Send rarely needs to be redirected in its sequencing, and can be extremely effective in identifying and moving the blocks needed by the processor in a timely fashion.
The 6th data column is the RPKI with PS, and a processor with a 32KB L1i and a 512-entry primary BTB.
%(details in section \ref{sec:Eval}).
%With PS a redirect happens when the fragment-level traversal of PS is different from the correct one identified by the processor's execution of the program. 
Note that the RPKI is an order of magnitude lower than
that with FDIP with an $\infty$ BTB, because the IPU is operating unperturbed by branches within a fragment, unlike FDIP which is redirected due to a branch direction misprediction within a fragment.
%\textcolor{red}{infinite BTB}.  
Consequently PS can "stay on track" much better than FDIP, and consequently achieve a very low fetch CWKI (10th column).  

%, even with a small BTB, and much better than FDIP with an $\infty$ BTB.

%\textcolor{red}{Consequently, it can efficiently supply the L1i, iTLB, and BTB, leading to low instruction wait cycles (mostly in the single digits), as reported in the 6th column for a 32KB L1i and a 512-entry BTB.}
%\textcolor{red}{Notably, this results in a significant reduction in magnitude, even compared to FDIP with an infinite BTB.}

%\input{DifferencesFromPrefetching}

\section{High-Level Program Map and Instruction Presending }
\label{sec:Send}

\subsection {Overview}

\begin{figure*}[h]
  \includegraphics[width=1.0\textwidth]{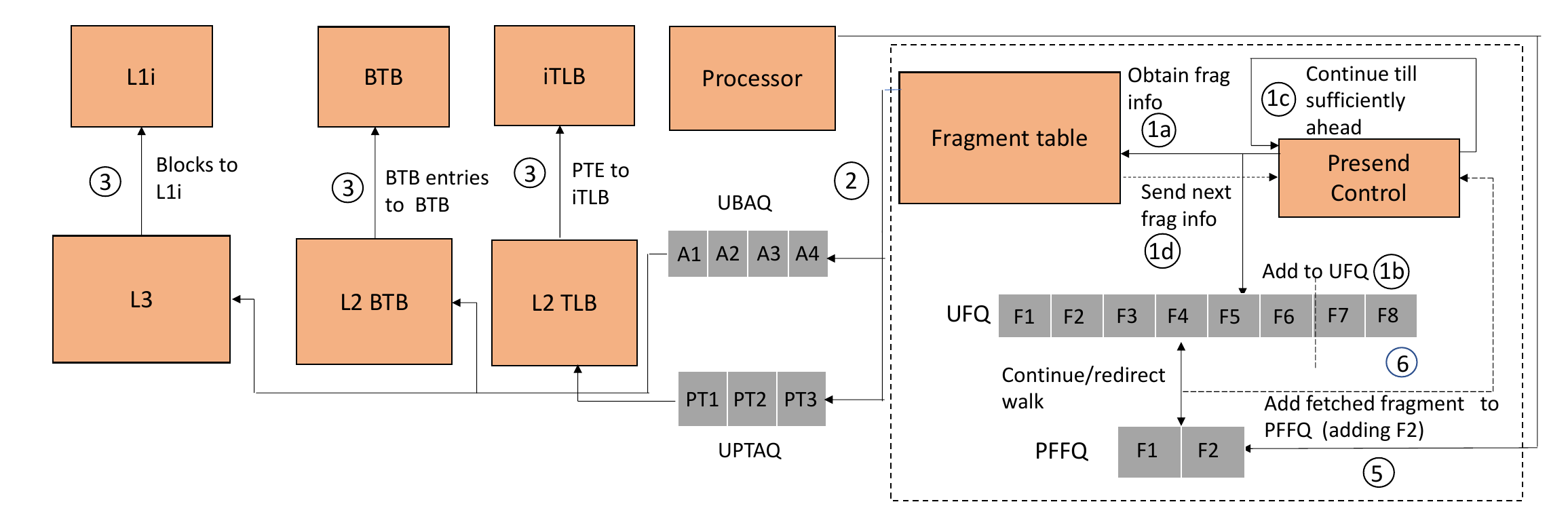}
	\caption
	{Overview of the Instruction Presending Unit (IPU)}
  \label{fig:send-mechanism}
\end{figure*}

%\begin{figure}[h]
%  \includegraphics[width=0.5\textwidth]{paper_diagrams_2/presend_cfl_with_forkpoints.pdf}
%	\caption
%	{Instruction Presending Unit (IPU) control flow prediction beyond fork points}
%  \label{fig:send-mechanism-2}
%\end{figure}

%\begin{figure}[h]
%  \includegraphics[width=0.4\textwidth]{paper_diagrams_2/program_snippet.pdf}
%	\caption[Functions (callsites), fragments and flow of control across fragments]%
%  {Program Structure and control flow across fragments \small {\normalfont Figure shows a sample program snippet and control flow across different function fragments}}
%  \label{fig:program-snippet}
%\end{figure}

%Motivating data in the previous section suggested that there are a small (few thousands) of program fragments and often
%there is only a single (next) fragment to which control flows after a given fragment.

The \textit{Instruction Presending Unit (IPU)}, works with a \textit{high-level program map}, or simply \textit{program map}, corresponding to a program's call graph.
Each step, or \textit{fragment} in the program map corresponds to a node in the call graph, and the transitions to the next fragment(s) correspond to edges.
The program map tracks the blocks in the fragment and the identifier(s) of the next fragment(s).
The IPU, shown in Figure \ref{fig:send-mechanism}, identifies the blocks in the fragment, and the identity of the successor fragment(s).
%(i.e., high-level sequencing).
It then accesses the identified blocks from the L3, if needed, arranges for them to be sent to the L1i/processor, and repeats the process with the next fragment.

The processor sends information about which fragment it is processing to the IPU
which uses it to track where the processor is, take corrective action (redirect) if needed,
and stay sufficiently ahead of the processor.
Along with code blocks, the IPU also sends iTLB and BTB entries 
from a secondary TLB/BTB to the primary TLB/BTB. 

In this paper, we construct the program map dynamically
(section \ref{Sec:FragmentConstruct}) and maintain it in hardware tables (section \ref{Sec:FragmentRepresent}).  We describe the IPU's operation in traversing the program map, 
block identification, and block movement collectively, as they are tightly coupled.
We first describe the sending of code blocks and then BTB and iTLB entries.

\subsection {High-level Program Map Representation and Creation}
\label{Sec:FragmentRepresent}

\begin{figure*}[h]
  \includegraphics[width=1.0\textwidth]{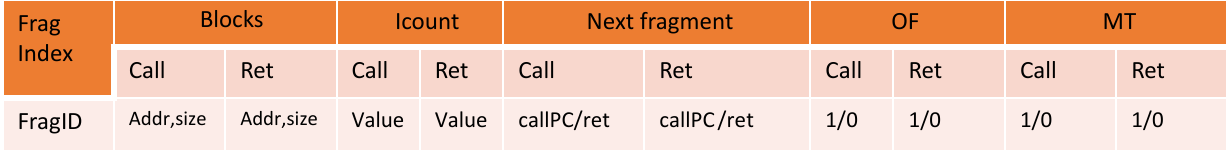}
	\caption[Fragment Table]%
	{Entry of a Fragment Table}
  \label{fig:frag-table}
\end{figure*}

\subsubsection {Main Fragment Table}

The main structure for representing fragments in the program map is a \textit{Fragment Table (FT)}.
A fragment starts at the target of a call or return instruction.
The PC of a call instruction is a natural identifier for the fragment
that starts with the first instruction of the called function.
Further, when that called function returns, the return target is the continuation
code from the next PC.
We combine both these related fragments into a single FT entry:
(i) a \textit{call fragment (Call)} and (ii) a \textit{return fragment (Ret)}.
For each fragment, the FT entry, shown in Figure \ref{fig:frag-table}, maintains:
(a) a set of blocks that are contiguous in the memory address space, or a \textit{region} of memory, 
(b) a count of the expected number of dynamic instructions in the fragment,
(or alternately the number of cycles to execute the code in that fragment),
(c) the identifier for a (single) \textit{next fragment}, or \textit{fragment target}\footnote{
We will use the terms next fragment identifier and target interchangeably.}.
The information for (a) is easily maintained as the address of the first code block (Addr) and a Size of the number of (contiguous) blocks in the region. The address of the first code block can be its full physical address. However, since PS only moves blocks from the L3 to the L1i, we can encode the address as a pointer to the block frame in the L3. The complete address can then be generated when the first block is read from the L3 using the tags. 
There is also a \textit{Multi Target (MT)} bit and an \textit{Overflow Regions (OF)} bit for each fragment, whose use we see next.

%\begin{figure}[h]
%  \includegraphics[width=0.5\textwidth]{paper_diagrams_2/dual_tgt_buf.pdf}
%	\caption[Dual Target Table]%
%	{Shows an example of an encoding in the dual-target buffer table}
%  \label{fig:multi-tgt-table}
%\end{figure}

\subsubsection{Dual Target Table (DTT)}

The data in Table \ref{fig:sig-callgraph-walk} suggested that mostly there is only a single target
per fragment, but often there are 2, and sometimes $>2$, targets per fragment.
To avoid increasing the size of the main FT, where we maintain only a single target,
we maintain additional targets in other tables.
A \textit{Dual Target Table (DTT)} maintains a second target fragment.
%, an entry of which is shown in Figure \ref{fig:multi-tgt-table} 
Additionally, we have observed that it is useful to maintain \textit{aging bits}
when there are multiple targets so that the additional (inactive) target
need not be pursued (section \ref{Sec:MultipleFrag}).
These aging bits for the two paths (one in the FT and the other in the DTT)
are maintained in the DTT for both paths\footnote{Since most fragments in the FT have only a single next fragment there is no need for aging bits for them.}. 

Like the DTT, information for more than two paths can be maintained in a \textit{Multi Target Table (MTT)}.
For many of the benchmarks we consider, maintaining two targets for a fragment is adequate though
for some benchmarks maintaining information about multiple targets can be beneficial.

%\begin{figure}[h]
%  \includegraphics[width=0.4\textwidth]{paper_diagrams_2/of_table_entry.pdf}
%	\caption[Overflow Table Entry]%
%	{Shows an example of an encoding in the overflow table}
%  \label{fig:of-table-entry}
%\end{figure}

\subsubsection{Overflow Regions Table (ORT)}

For many fragments, there is a single (contiguous block) region that can be represented in the main FT.
However, several fragments have multiple non-contiguous blocks.
%arising from more complex local control flow. 
%To accommodate and encapsulate the complex local control flow in such fragments,
For such fragments, we maintain the additional
non-contiguous regions in an \textit{Overflow Regions Table (ORT)}.
%an entry of which is shown in Figure \ref{fig:of-table-entry}.
Like the FT, an ORT entry has the starting block address and size to represent a region.

To accommodate the skewed distribution of the number of code regions in a fragment,
we have found it useful to have multiple ORTs with different numbers of additional code regions.
For example, ORT-2, ORT-4, ORT-16, where each entry can represent 2, 4, or 16 
code regions in a fragment.

%\begin{figure}[h]
%  \includegraphics[width=0.5\textwidth]{paper_diagrams_3/multi_tgt_code_frag.pdf}
%	\caption[Code Fragment (control flow to two fragments)]%
%	{Shows an example of a code fragment where control can flow from a single fragment to two fragments}
%  \label{fig:multi-tgt-frag-code}
%\end{figure}

\subsection {High-level Program Map Construction} 
\label{Sec:FragmentConstruct}

\begin{figure*}[h]
  \includegraphics[width=0.97\textwidth]{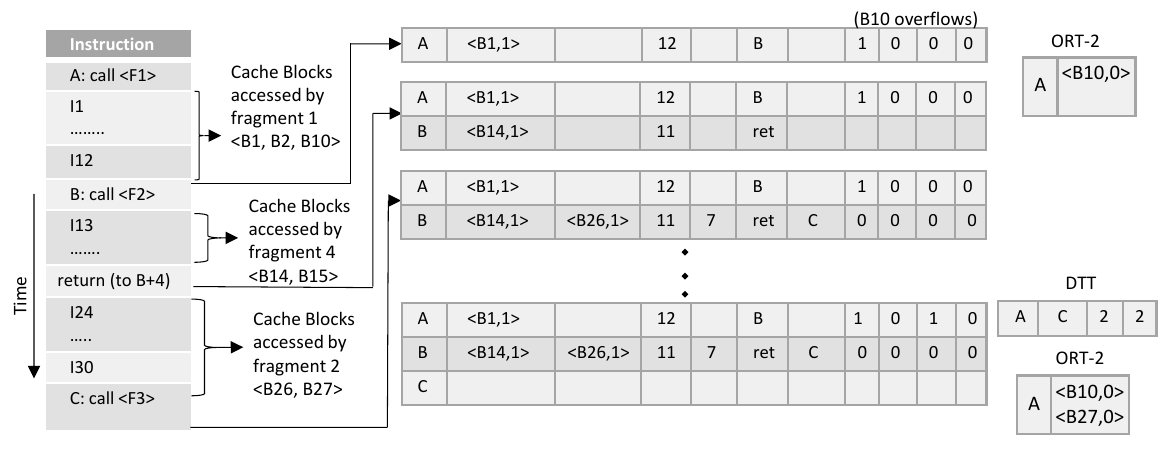}
	\caption[Fragment Table construction]%
	{Fragment Table Construction Example}
  \label{fig:cgraph-ds-constr}
\end{figure*}

We describe how FT (and DTT/ORT) entries are populated in Figure \ref{fig:cgraph-ds-constr}
for the execution of the program snippet of Figure \ref{fig:program-snippet}. 
When the processor retires a call (at PC A), it starts a new fragment \textit{frag1} (call fragment with ID A). Say the if condition in function F1 is true and thus the call at PC B is encountered.
At this point the fragment \textit{frag1} is terminated, a new fragment \textit{frag4} (call fragment with ID B) starts, and thus, the next fragment for \textit{frag1} (call fragment with ID A) is \textit{frag4} (call fragment with ID B). The instruction blocks accessed by this fragment are B1, B2, and B10. The processor also tracks the dynamic number of instructions executed (say 12)  between the call at PC A and the call at PC B\footnote{
Alternatively, we could track the time between when the two different calls reach a processing (e.g., decode) stage.}.

As the processor transitions from retiring instructions in fragment \textit{frag1} to fragment \textit{frag4}, the processor populates an entry in the FT for fragment \textit{frag1} with the gathered information. The fragment has two contiguous blocks (B1, B2), and thus the main FT entry has \textlangle B1,1\textrangle indicating that B1 is the first block
in the fragment, and there is one additional block.  Since B10 is not contiguous, it is put in an ORT-2 in the example, which tracks up to 2 overflow regions, and the ORT bit set; the ORT-2 entry has \textlangle B10,0\textrangle.  Further, the Count is set (to 12).

Continuing further, after the call at PC B, blocks B14 and B15 are encountered, followed by a return. Thus, when the return instruction retires, an entry for fragment \textit{frag4} (with ID B) is created with \textlangle B14,1\textrangle, with a Count (of 11) and 
the next fragment being set to return.
The return then returns to the code block at PC B+4, thus starting fragment \textit{frag2} (the return (ret) fragment associated with B).
Say code blocks B26 and B27 are accessed, followed by the call at PC C.
On retiring the call instruction at PC C, the entry for fragment \textit{frag2} is created with
\textlangle B26,1\textrangle, a Count (of 7) and the next fragment being set to C.

At some point later in the program execution the if condition is false, and after the call at PC A
(execution of \textit{frag1}) execution flows to the call at PC C. 
This causes a \textit{redirect} of the IPU (below) because, as per the current FT entries,
the IPU expected the next fragment following fragment \textit{frag1} to be fragment \textit{frag4}.
At this point, the next fragment information for \textit{frag1} (at ID A) is updated by creating an
%Since there is already an FT entry for ID A with the next fragment ID as B,
entry in the DTT for ID A with the next fragment ID as C,
and the MT bit in the FT entry is set.
The blocks accessed by the fragment are also updated for the additional blocks before the call to C which in this case is block B27, that is updated in the ORT-2 with \textlangle B27,0\textrangle.
%if any of the additional blocks miss in the L1i.
Further, aging bits for the two paths (2,2) are also set (in the DTT).

At the start of program execution, the FT is empty.
For each new fragment retired the IPU is redirected, and an FT entry for that fragment is created.
After creation, an entry is updated only on L1i misses (or IPU redirects)
during the fragment's execution. 
%ot every time a fragment is retired.
Otherwise, the FT information used by the IPU to supply the blocks is correct, and no updates are required. 
We will see that empirically there are very few L1i misses and IPU redirects,
thus the FT (and associated tables) are quickly populated and then heavily reused.

%With this, entries in the FT/DTT/ORT are created and updated when there

\begin{algorithm}[htb]
\caption{Operation of the IPU}
\label{alg:SendingAlgorithm}
\SetKwInput{KwInput}{Input}                % Set the Input

\SetKwInput{KwOutput}{Output}              % set the Output
\DontPrintSemicolon

%
% 
%
% Set Function Names
%
%  \SetKwFunction{compaqt}{{Get\_Compressed\_Pulse}}
%
%
% 
%
%  \SetKwProg{Fn}{Function}{:}{}
%
%  \Fn{\compaqt{$W_{\text{in}}$, $\varepsilon$}}
%
%  {
%
%	current\_fragment = first\_fragment
%	
%	current\_lookahead = 0

	%FragInfo = empty
	%\While {true}{

	Next\_Frag = Starting Fragment from Processor

	\While{not sufficiently ahead (Step 1c)}{
		\uIf{Next\_Frag = Call}{

			Access FT (Step 1a)	
		
			Push Return Fragment in IPUS

			FragInfo = FragInfo from FT (Step 1d)
	
			Update Next\_Frag from FragInfo (Step 1d)
	        
		}
		\uElse {
			FragInfo = Pop from IPUS

			Update Next\_Frag from FragInfo (Step 1d)
		}		
		
		Use FragInfo to update UFQ (Step 1b)

		Add Blocks from FT to UBAQ (Step 2)
	}
	
%	//Send Engine blocks waiting for processor cue.

%	Wait for Processor Cue
%
%	Processor reaches proc\_fragment (cue to the Send Mechanism). 
%	
%	\uIf{proc\_fragment in queue}{
%		
%		current\_lookahead $=$ current\_lookahead $-$ icount(proc\_fragment)
%	
%	}
%
%	\uElse{
%		
%		//Redirect
%		
%		current\_lookahead = 0
%		
%		current\_fragment = proc\_fragment
%	
%	}
	
%	}		
%

\end{algorithm}

\subsection{Operation of the IPU}
\label{Sec:OperationDetail}

\subsubsection{Single Next Fragment}
\label{Sec:SingleFrag}

%\vspace{-0.1in}
%\setlength{\textfloatsep}{0pt}
%\SetAlgoNoLine
%{

%\begin{algorithm}[htb]
%\caption{Operation of the IPU}
%\label{alg:SendingAlgorithm}
%\input{algorithms/send_algo_1.tex}
%\end{algorithm}

%}

%\begin{figure}[h]
%  \includegraphics[width=0.5\textwidth]{paper_diagrams_2/callstack_ops_1.pdf}
%	\caption{IPU Stack (IPUS) }
%  \label{fig:stack-ops}
%\end{figure}

%\begin{figure}[h]
%  \includegraphics[width=0.5\textwidth]{paper_diagrams_2/frag_supply_queue.pdf}
%	\caption{Upcoming Fragments Queue (UFQ)}
%  \label{fig:frag-supply-queue-1}
%\end{figure}

We start by explaining the IPU operation for fragments with a single target
in Algorithm \ref{alg:SendingAlgorithm}, which
indicates the steps taken in reference to the elements of Figure \ref{fig:send-mechanism}.
The IPU works with a fragment ID and
tries to stay sufficiently ahead of the processor.
If the fragment ID is a call, it is used to access the FT (step 1a) and obtain the 
call path fragment information (blocks in fragment, count, and next fragment ID).
Further, the fragment information for the return fragment (also obtained from the FT) is pushed onto a
\textit{IPU stack (IPUS)}.
%as shown in Figure \ref{fig:stack-ops}.
If the next fragment ID is a return, the fragment information is popped from the IPUS.

The fragment ID is recorded in an \textit{Upcoming Fragments Queue (UFQ)} (step 1b).
%as shown in Figure \ref{fig:frag-supply-queue-1}. 
The addresses of the blocks in that fragment
(from the FT/ORT)\footnote{
If a return fragment also has entries in the ORT, those regions are also pushed
onto the IPUS along with the information from the FT.}
are put into an \textit{Upcoming Block Addresses Queue (UBAQ)} (step 2).
The next fragment ID (step 1d) is used to repeat the process
and keep the IPU sufficiently ahead of the processor (step 1c).
For blocks in the UBAQ, a decision is made (section \ref{sec:BlockControl})
whether to read a block from the L3 and send it to the L1i (step 4).
%The blocks to be sent are accessed from the L3 from where they proceed to the L1i/processor (step 4).
%The blocks that are to be sent are accessed from the L2 cache and put into a
%\textit{Presend Block Queue (PBQ)} (step 3) from where they proceed to the L1 cache/processor (step 4).

Overall, the IPU's operation is very simple: read a table entry, send parts of the entry to some buffers (the sending part),
and use part of the entry to get the index of the next entry or two (the high-level program map traversal part) and repeat.

\subsubsection {Processor-IPU Synchronization }

The IPU tries to stay on the correct execution path,
and sufficiently ahead of the processor.
For this, it maintains a \textit{Processor Fragment Fetch Queue (PFFQ)} to monitor
the processor's progress.
%where the processor is in its execution of the program.
As the processor fetches a fragment, it inserts the fragment ID
(call PC or return address) into the PFFQ (step 5).
The IPU compares the PFFQ and UFQ entries to detect when it is on
a different path from the processor and redirects
in case of divergence.
%the IPU is redirected.
Redirects happen when there is complex control flow behavior at the call graph level
that the FTT/DTT have not captured.
One example, in our evaluated implementation, is that we track only two target fragments,
and the processor goes down an untracked path.
Another example is when there are unmatched calls and returns.
When redirected, the IPU starts on a new path with the
latest fragment ID sent by the processor, discarding other activity.

%\footnote{Divergence could happen for
%a variety of reasons, including call.

Knowing which fragment the processor has reached,
the IPU can keep sufficiently ahead
by using the Count information of the (remaining) fragments in the UFQ.

%We could also reduce the frequency of communication, e.g., once for every N fragments,
%to trade off the amount of processor-IPU communication and the latency of
%correcting the IPU when it is proceeding incorrectly.

\subsubsection{Multiple Next Fragments}
\label{Sec:MultipleFrag}

When there are multiple potential next fragments (the MT bit is set), 
the IPU has a choice of what to do:
%for the high-level sequencing: 
proceed down only one path or down multiple paths?
The aging bits in the DTT/MTT are used to make this decision.
To proceed down multiple paths, the IPU has separate tracks, 
one for each path.
%each proceeding down a single path.
A copy of the IPUS is created at a fork point and each path is given its own IPUS.
The fragment information for the second path is obtained from the DTT and each track of the IPU proceeds
down a single path.
Eventually, one of the paths is determined to be the correct path, using the processor fragment IDs
in the PFFQ.
The incorrect paths are discarded and the IPU proceeds down the correct path.
Aging bits are updated positively for the correct path
and negatively for the incorrect one.
Our experimental results suggest benefits in proceeding down two paths for almost all our benchmarks,
and in proceeding down multiple paths for some.

\subsection {Additional Program Control Constructs}
\label{Sec:AdditionalConstructs}

%\begin{figure}[h]
%  \includegraphics[width=0.5\textwidth]{paper_diagrams_3/indirect_code_frag.pdf}
%	\caption[Code Fragment showing indirect call]%
%	{Shows an example of code fragment with indirect call having multiple active targets}
%  \label{fig:ind-frag-code}
%\end{figure}

%\begin{figure}[h]
%  \includegraphics[width=0.5\textwidth]{paper_diagrams_3/indirect_buffer.pdf}
%	\caption[Fragment Table with indirect callsites]%
%	{Shows an example of an encoding in the Fragment Table in the presence of indirect callsites with multiple targets}
%  \label{fig:ind-frag-table-encoding}
%\end{figure}

\subsubsection {Indirect calls with multiple targets}

For direct calls, there is a unique fragment associated with the call, and thus the PC of the call is an adequate fragment identifier.
For indirect calls, there could be multiple targets of the call.
To distinguish between the different targets, we use a hash of the call PC and the target PC
as the fragment identifier.
Note that the IPU works with a fragment identifier as created; it does not know
if the call was indirect or not.
%Figure \ref{fig:ind-frag-code} we see that the call at A has targets of T1 (<frag1>), T2 (<frag2>) and T3 (<frag3>) and thus
%<frag1>, <frag2>, and <frag3> are A hash T1, T2, T3, respectively to identify them in the FT.

%\begin{figure}[h]
%  \includegraphics[width=0.5\textwidth]{paper_diagrams_3/recursion_code_frag.pdf}
%	\caption[Code Fragment showing recursion]%
%	{Shows an example of fragment control flow with recursion and callstack contents}
%  \label{fig:recursion-frag-code}
%\end{figure}

%\begin{figure}[h]
%  \includegraphics[width=0.5\textwidth]{paper_diagrams_3/recursion_buffer_tbl.pdf}
%	\caption[Fragment Table encoding with recursion]%
%	{Shows an example of an encoding in the Fragment Table in the presence of recursion}
%  \label{fig:recursion-table}
%\end{figure}

\subsubsection {Loops and Recursion}

For loops and recursion the same set of fragments are executed repeatedly, and eventually execution proceeds
to the fragment at the continuation of the loop or the recursive call.
The IPU need not send blocks from the loop fragments repeatedly as they will likely be in the L1i 
after the first sending.
After the loop/recursion exit, the IPU should be sufficiently ahead to avoid misses on the continuation path.
Accordingly, at a loop/recursion, the IPU proceeds along two paths, one along the looping/recursion path
(which can be of an indeterminate length), and the other along the continuation path.
For the former path, the IPU does little after the initial sends other than
monitoring the fragment IDs along this path sent via the PFFQ.
Along the latter path, it attempts to stay a certain distance ahead of the processor.
Eventually, when the IPU sees a fragment ID from the latter path in the PFFQ, it discards the former path.

%We also note here that such static points are very few for most of the server applications, which require such renaming. 

%\begin{figure}[h]
%  \includegraphics[width=0.5\textwidth]{paper_diagrams_3/loop_code_frag.pdf}
%	\caption[Code Fragment showing loop control]%
%	{Shows an example of fragment control flow with loops}
% \label{fig:loop-frag-code}
%\end{figure}

%\begin{figure}[h]
%  \includegraphics[width=0.5\textwidth]{paper_diagrams_3/loop_table.pdf}
%	\caption[Fragment Table encoding with loops]%
%	{Shows an example of an encoding in the fragment predictor table in the presence of loops}
%  \label{fig:loop-table}
%\end{figure}

\subsection{Fragment Table Set Associativity}
\label{sec:FragAssoc}

The FT is set associative as the associativity helps reduce the number of conflicts on fragment creation.
%On a fragment creation, the FT is accessed to determine an entry;
%the associativity helps to reduce the number of conflicts.
However, the IPU accesses the FT using the next fragment ID, not an arbitrary fragment ID.
If the next fragment ID is maintained as an index into the FT,
most of the IPU's FT accesses can be direct, \footnote{The
only time a set associative access would be needed is when the IPU is redirected and restarts with a new fragment ID.}
resulting in increased speed and reduced FT access energy.
Likewise, accesses to the DTT and ORT can also be made directly rather than set associatively,
as they merely maintain additional information for a fragment in the FT.

\subsection{Enhancements for an iTLB/BTB}
\label{sec:ITLBEnhancements}

%To also send iTLB entries, the FT entry adds a pointer to an L2 TLB entry with each region.
%When sending code blocks from a region (step 2), the pointer is used to access the L2 TLB and put the corresponding
%entries in an \textit{Upcoming Page Table Addresses Queue (UPTAQ)} (step 2), as shown in Figure \ref{fig:send-mechanism}. 

To send iTLB entries, the FT entry (and ORT entry) adds a pointer to an L2 TLB entry with each region.
When sending code blocks from a region (step 2), the pointer is used to access the L2 TLB and put the corresponding
entries in an \textit{Upcoming Page Table Addresses Queue (UPTAQ)} (step 2), as shown in Figure \ref{fig:send-mechanism}. 
From there, they proceed to the iTLB.
As in the case of cache blocks, appropriate presence checks to reduce unnecessary movement can be made, if needed.
Eviction of an L2 TLB entry will result in a subsequent iTLB miss and a consequent updating of the FT entry.

To also send BTB entries, the virtual page numbers are extracted from the tag bits of the corresponding iTLB entries 
and used along with the block address in the UBAQ to generate addresses for the L2 BTB.
The corresponding entries accessed (taking multiple accesses as needed)
and sent, as also shown in Figure \ref{fig:send-mechanism}, making presence checks as needed.

%\subsubsection {Future Knowledge for L1 Cache Replacement}
%\label{sec:FutureKnowledge}

%The block addresses in the UBAQ are the blocks that are expected

%By staying ahead of the processor, the block addresses in the UBAQ are
%essentially the blocks that the processor
%is going to be \textit{referencing in the near future}.

\subsection {Location of the IPU}
\label{sec:IPULocation}

Since the IPU is intended to operate autonomously,
it can be placed at different points in the processing pipeline.
One option is to place the IPU alongside the L1i cache.
In this case, the blocks in the UBAQ that are accessed from the L3 and sent to the
L1i and can be "pulled" from the L3 just as if they were L1i misses using existing miss paths.
Another option is for the IPU to be near the L3, where it could potentially be shared by different cores running the same program.
In this case, the IPU would access the needed blocks from L3 and "push" them to L1i.
%While we have the processor populating the FT with its retirement stream,
%the FT could also be constructed in the L3 from the block references that reach the L3 between fragment IDs (to be explored later). Or in software.

%Regardless of its placement, primed with an FT, the IPU only needs to know the processor's fetch fragment IDs to operate and correct
%its high-level sequencing.  This communication isn't significant, as we shall quantify.
%It can be reduced by communicating once every N fragments,
%trading off the amount of communication with the latency of redirection when IPU is proceeding incorrectly.

\subsection {Processor Wrong-Path Execution}

If the processor inserts fragment IDs into the PFFQ at fetch, then there is the possibility of incorrect IDs being inserted when the processor is fetching on a wrong branch path, leading to unnecessary IPU redirects.
Due to the high-level program map that the IPU is traversing, this is a very infrequent occurrence as we shall quantify in Table \ref{fig:rpki}.
First, for many fragments, there is a single successor fragment, and regardless of the branch outcomes in the fragment, the next fragment inserted into the PFFQ is the same.  (Similar observations were made by other studies exploiting Control Independence \cite{sohi1995multiscalar,cher2001skipper}.)
Second, for fragments with multiple successors, we record two successors in the FT/DTT, and the IPU proceeds down two paths. Thus, the IPU will only be (incorrectly) redirected if the processor's execution on a wrong branch path leads to a third, different fragment.

\subsection {Block Presending Decision Control}
\label{sec:BlockControl}

%By sequencing at the fragment level, without being perturbed by dynamic fragment-internal branches (an average of over 4 in our experiments), the IPU can do a better job of keeping on track, with fewer redirections, than approaches that may be redirected by every (mispredicted) dynamic branch.
Since the IPU dumps large chunks of block addresses into the UBAQ
there is the potential for a significant increase in the L3 accesses and sending of useless blocks.
%that are not useful. 
%which could also lead to pollution in the L1i.
Accordingly, a decision needs to be made about which blocks in the UBAQ
\textit{should be accessed from the L3 to be sent to the L1i}.

%We address this next.

Blocks could be wastefully sent because: (i) the IPU proceeds down two paths, as in section \ref{Sec:MultipleFrag},
with blocks from one path not being used,
(ii) blocks within a fragment are not going to be used based upon the branch outcomes within the fragment, and
(iii) a block is already in the L1i.

For the large code footprint applications that we consider,
most of the time there is only a single successor fragment (and path).
%illustrative examples of which were shown in Table \ref{fig:sig-callgraph-walk}.
An infrequent path is also not followed,
as per the aging bits in section \ref{Sec:MultipleFrag}.
Further, two paths are followed, when needed, only for a short period.
Collectively this results in only a small overhead,
as we quantify in section \ref{L3Traffic}.

%There are two reasons why the sending of a block will be wasteful:
%(i) the block is already in the L1i, and (ii) the block is unlikely to be referenced by the processor
%because it is from a cold region of code.

Large server programs have regions of cold code \cite{ayers2019asmdb}.
%Untouched cold code regions aren't brought into the L3.
%However, some regions may have been touched,
%for which entries are created in the FT/DTT/ORT, but are rarely accessed later.
This can result in blocks within a fragment in the FT/DTT/ORT
becoming cold while other blocks continue to stay hot.
To track the \textit{temperature} of code blocks we can associate a set of \textit{temperature bits (TBs)} with a block frame in the L3.
%However, temperature bits with an L3 block frame would mean probing the L3 cache
%tags to determine their value, and this is something that we would like to avoid.
Rather than associate them with the L3 tags, which would require an L3 tag access,
we use a \textit{Block Temperature Table (BTT)}.
The BTT is a linear array indexed with bits of the block address
(e.g., 11 bits for a 2K entry table).  An entry in the BTT is an n-bit (e.g., 3-bit) temperature value.
The temperature bits in a BTT entry are set to a high temperature when a block mapping to that BTT entry is sent.
An additional \textit{accessed} bit is added to each block frame in the L1i.
The accessed bit is reset when a block is placed in the L1i, and set when it is accessed.
%When a block is sent to the L1i, the accessed bit is reset.  Then if it is accessed before being evicted,
%the accessed bit gets set.  
If a block is evicted from the L1i with its accessed bit reset,
the temperature bits in the corresponding BTT entry are decremented.
The BTT entry is checked for blocks in the UBAQ and only blocks with an adequate temperature are sent to the L1 cache via the PBQ.
Because multiple blocks/frames map to the same BTT entry, they end up sharing and updating
the temperature bits collectively.  We have found this aliasing to not
be a major issue.
%concern as the temperature information is merely advisory.

%The approach to not sending blocks that might already be in the L1i depends upon the placement of the IPU  (section \ref{sec:IPULocation}). If the IPU, is alongside the L1i, it could probe a copy of the L1i tags to make this determination. For an IPU placement elsewhere, e.g., near the L3, probing L1i tags is not practical.

To check for block presence in the L1i, rather than probe the L1i tags, we
a pseudo-inclusion bit per block in
%to know whether or not a block is present in the L1i. 
a \textit{Pseudo Inclusion Bit Table (PIT)}, analogous to the BTT.
The PIT is accessed with the low-order bits of the block number, and entries updated when a block is moved to,
or replaced from, the L1i.  We have observed that
using a PIT 
%to control block movement 
provides similar results as probing the L1i cache tags.

\section{Prefetching Schemes and Related Work}
\label{sec:Rel-work}

%Before evaluating presending we discuss some prefetching schemes to understand the differences and
%why presending can be more effective.

%Popular instruction prefetching schemes, exemplified by

\textit{Fetch directed instruction prefetch (FDIP)} \cite{reinman1999fetch,ishii2021re} is a widely used and
very effective scheme.  It works to establish the upcoming instruction stream by determining branch outcomes,
using \textit{branch direction predictors} and BTBs, and prefetching the corresponding cache blocks.
Many proposed prefetching schemes improve on FDIP by improving its ability to
determine branch outcomes, typically by improving the primary BTB capabilities,
and improving the identification of blocks to prefetch.  
Two recent proposals are Confluence \cite{kaynak2015confluence} and \textit{Shotgun} \cite{kumar2018blasting}. 
Confluence proposes to unify the metadata to prefetch into both L1i and the primary BTB.
Shotgun, which is more effective, goes further and proposes a novel primary BTB design that has multiple components and is more effective than a standard BTB of a similar size.
An unconditional BTB (UBTB), based on high-level control flow, maintains a block footprint which is
used to trigger prefetching into the L1i and into a primary conditional BTB (CBTB).
The structure of the UBTB is like the main FT that we employ;
information that we maintain in the ORTs is spread across multiple UBTB entries. 
All of these schemes still rely on a branch predictor, and the primary BTB,
to sequence every branch in order to anticipate the upcoming stream of instructions.
Their ability to get ahead is influenced by the branch predictor/BTB accuracy and effectiveness.

%which we use for our comparison further attempts to reorganize the BTB based on high level control flow into a 
%unconditional BTB with a block footprint and using this to prefill the conditional BTB just-in-time, 
%and using it to trigger block prefetches.

%Confluence \cite{kaynak2015confluence} is a scheme that attempts to prefill BTBs using same metadata used for prefetching blocks of instruction. Similar to Confluence, Send also inserts all BTB entries from a block being sent to the L1i into the BTB, though using a shadow representation and not a recorded miss stream.  

Another class of instruction prefetching schemes, of which \textit{Entangling Prefetch} \cite{ros2021cost}
is the most effective, learn about blocks that will miss and tie them to instruction stream events (e.g., instruction cache accesses),
and prefetch when those events are triggered
\cite{srinivasan2001branch,ferdman2008temporal,ferdman2011proactive,kolli2013rdip,ansari2020divide,ansari2021mana,ros2021cost}. 
Such prefetchers, while not relying explicitly on BTBs, etc., are still influenced by the correctness of the instruction stream which impacts the triggering events.
Moreover, by themselves, they don't reduce the reliance of the fetch mechanism on a large primary BTB;
this requires a separate BTB prefetching mechanism or more complex large BTB designs \cite{bonanno2013two}. \textit{Return-directed instruction prefetching (RDIP)} \cite{kolli2013rdip} and \textit{call graph prefetching (CGP)}  \cite{annavaram2003call} also seek to exploit a higher level of program control flow, 
though in a very controlled manner, staying one fragment ahead and prefetching blocks from the next node in the callgraph.
%tied to processor return address stack acitivity. 
They are not as effective as recent prefetching techniques \cite{ros2021cost}.

%such as return address stack activity, 

%A recent example of an advanced prefetching technique (in the second category), which we used for comparison in this paper,
%\textit{Entangling Prefetch} \cite{ros2021cost} correlates cache misses to prior access events,
%initiating a (timely) prefetch when the prior access event occurs.

Presending, which traverses a higher (call graph) level map of the program,
operates autonomously and does not rely on branch predictors or BTBs, or other processor microarchitectural events such as cache accesses and cache misses.
Indeed we use the term presending to describe the technique, rather than prefetching since generally the operative actions of a prefetching scheme are influenced by processor microarchitectural events tied to fetching.  Presending's operative actions are determined solely by the high-level program map traversal, and not processor events.  It only relies on fragment IDs that the processor is executing to stay sufficiently ahead and on track.  Since presending is not redirected by branch mispredicts or BTB misses, it can better stay on track (which we quantify in Section \ref{Redirects}) for providing instruction blocks 
%(as opposed to discovering the precise instruction stream),
and associated iTLB and BTB entries.
The high-level program traversal is influenced by the control flow concepts in Multiscalar, which facilitated the creation of extremely large instruction windows agnostic of local branch mispredictions \cite{pnevmatikatos1993control,sohi1995multiscalar}. 
Presending in effect provides an appearance of large primary memory hierarchy structures
in the processor while having much smaller structures.
Predictor virtualization \cite{burcea2008predictor,liu2024avm} has similar objectives, but uses very different techniques.
%Further it does require large primary BTBs (or different designs) and can move needed entries into a small processor primary BTB just in time.

%Maybe mentione inspector-executor, multiscalar, others?

\section{Evaluation}
\label{sec:Eval}

\begin{table}[]
\centering

\caption{Simulated Machine Parameters}

\scalebox{1.0}{
\begin{tabular}{ll}
\hline
	\multicolumn{2}{c}{\textbf{Processor Decoupled Front-end}}                \\ \hline
	Width                & uA1 - 6 instr; uA2 - 8 instr                            \\
	Fetch/Decode/Dispatch queue          &  192/60/60 instr                                \\
	%Decode queue         & 60 entry                                \\
	%Dispatch queue       & 60 entry                                \\
BTB/Target Cache/RAS & 8K/4K/64 entries                                \\
%Target cache         & 4K entries                                \\
%Return Address stack & 64 entries                                \\
Branch penalty       & 2 cycles (decode stage)                   \\
Branch Predictor     & Hashed perceptron                         \\ \hline
	\multicolumn{2}{c}{\textbf{Processor Back-end}}                         \\ \hline
Execute/Retire width        & uA1 - 4/5 instr; uA2 - 10/8 instr                           \\
%Retire width         & uA1 - 5 instr; uA2 - 8 instr                            \\
Re-order buffer      & uA1 - 352 entries; uA2 - 1000 entries                               \\
	Load, store queue    &  128, 72 entries                         \\ 
	Load, store queue (uA2)    & 300, 300 entries                         \\ \hline
	\multicolumn{2}{c}{\textbf{Memory hierarchy}}                           \\ \hline
L1i cache            & 32KB, 8-way, 4 hit cycles, 16 MSHR  \\
L1 iTLB              & 64-entry, 4-way, 1 hit cycle              \\
L1-D cache           & 48KB, 12-way, 5 hit cycles, next-line;     \\
L1-D cache (uA2)           & Perfect, 1 hit cycle, next-line     \\
L2 cache             & 512 KB, 8-way, 10 hit cycles, spp-dev     \\
	L2 BTB &     16K-entry, 8-way, 8 hit cycles \\
	L2 TLB &     2K-entry, 8-way, 8 hit cycles \\
L3 cache             & 2MB, 16-way, 20 hit cycles, no pref \\
DRAM                 & 4 GB, one 8B channel, 1600 MT/s      
\end{tabular}
}
\label{fig:sim-parameters}

 %\vspace{-10pt}
\end{table}

\begin{table}[]
\centering
	\caption{Table Storage Requirements}

\scalebox{1.0}{
\begin{tabular}{|l|l|l|l|}
\hline
Structure                    & Entries      & Entry Size                                                                                 & Storage \\ \hline
	 FT         & 4096 & 12.375B &   50688B   \\ \hline
	DTT  & 256 & 3.375B  &   864B    \\ \hline
 BTT  & 2048 & 0.375B                                                                                          &   768B 
 \\ \hline
		ORT-2  & 1024 & 6.375B     &   6528B    \\ \hline
		ORT-4  & 256 & 11.375B     &   2912B    \\ \hline
		ORT-16  & 128 & 41.375B                                                                                          &   5296B    \\ \hline
	%	BTT  & 2048 & 0.375B                                                                                          &   768B    \\ \hline
\end{tabular}
}

	%\vspace{-10pt}
\label{fig:presend-storage-overhead}

\end{table}

\subsection {Simulation Details}
\label{SimDetails}

We evaluate the instruction presending mechanism (abbreviated \textit{PS}) using the ChampSim simulator \cite{gober2022championship}
and 100 server benchmark traces provided by Qualcomm Datacenter Technologies.
These benchmarks and simulator have been used in multiple recent papers on prefetching 
%\cite{chaconsw},\cite{asheim2023storage},
%in CVP-1 \cite{cvp1}, 
\cite{chaconsw, asheim2023storage,ros2021cost}. In this work, we simulate the ARM ISA.
We group the benchmarks into 5 bins, based upon the base MPKI with a default L1i size.
Bins I, II, III, IV, and V correspond to base MPKIs of 10-15, 15-30, 30-45, 45-60, and $>$ 60, and have 16, 45, 6, 23, 10 benchmarks, respectively.

%\footnote{The other classes of applications have very low icache MPKIs} 

%used in CVP-1 \cite{cvp1}, which have been used in multiple other papers on prefetching \cite{gober2020temporal,gupta2020runjump,ansari2021mana,ros2020entangling,ros2021cost}.
%In particular, we use the same benchmark traces (all the Server traces) as those use by the work on \textit{Entangling Prefetch (EP)} \cite{ros2020entangling,ros2021cost}.

The baseline configuration (or uA1) described in Table \ref{fig:sim-parameters}, 
implements a decoupled front-end modeling FDIP \cite{reinman1999fetch}, with
a BTB, a Target Cache to predict the target of indirect branches \cite{chang1997target} 
and a return address stack (RAS).
%, similar to what was used in \cite{ros2021cost},
%as recent work \cite{ishii2021re} emphasizes use of an FDIP baseline.
The simulated FDIP design has a single-level, 1-cycle BTB, which is more aggressive
than designs that employ a small, 1-cycle L1 BTB corrected by a larger L2 BTB with longer access latency.
%Prefetches issued by the Fetch-Directed Prefetching engine are considered demand accesses.   
Table \ref{fig:sim-parameters} also describes uA2, a microarchitecture with wider fetch (8-wide)
and a more aggressive backend (large ROB/SQ/LQ and a perfect D-cache).
%uA2 shares most microarchitectural parameters with uA1, with the differences highlighted in the Table.
We simulate a 16K-entry L2 BTB and a 2K-entry L2 TLB from where entries are sent to the corresponding small-sized L1 structures. 
Since these structures aren't used to "correct" a smaller L1 structure (unlike an L2 BTB for FDIP), but rather are to provide capacity,
they can be slow (8 cycles in our case).
Cache blocks are sent from the L3 to the L1i.

%Sent cache blocks/BTB entries/iTLB entries are directly installed in the respective structures,
%without additional buffering.

%We do not use any buffers (such as prefetch buffers) for our simulations, including for a smaller sized iTLB and BTB.
%All processor accesses are made only to L1i/iTLB/BTB. 

We compare against two of the most effective prefetching schemes.
The first is \textit{Shotgun (SG)}; we use a configuration with a 4K entry UBTB and a 1K entry CBTB,
for which the storage is about 64KB.
Smaller configurations weren't as effective for our benchmarks.
Further, the SG configuration that we simulate is very aggressive.
For PS, for a fragment with multiple code regions (due to an unconditional branch
from one region to another), the multiple regions reside in the ORTs and are processed in one step.
In SG, they reside in multiple UBTB entries, but our simulation treats these multiple UBTB accesses as a single access.
%Our simulated SG implementation also processes them in one step, simulating a single access for the multiple UBTB entries.
The second is \textit{Entangling Prefetch (EP)}, and we use a configuration with a
storage overhead of 77.44KB \cite{ros2021cost}. All schemes are implemented on top of the FDIP baseline in
Table \ref{fig:sim-parameters}. 
Though all schemes are implemented on top of FDIP similar to recent works \cite{ros2021cost,ishii2021re}, our experiments indicate that PS performance (not shown) remains comparable even when FDIP is not used as the baseline.

%We are mostly interested in evaluating the effectiveness of Send in providing instruction cache blocks
%to the L1 cache from L3 in the on-chip memory hierarchy.
%Our comparison are carried out with the best instruction prefetcher that we are aware of: 
%Entangling Prefetcher with 8K entries (EP), which takes up 77.44 KB of storage \cite{ros2021cost}.

%Say that Shotgun is very aggressive.  Describe how it is agressive --- goes past unconditional branch just like we have in our ORT in the same cycle.
%Multiple entries for discontiguous regions that we send in one cycle as if it were one entry.

%Say that we don't use a shadow cache.  Rather we just use the prefetch structure of FDIP and simply pass blocks from the whatever queue to the
%prefetch part.

For speedup, we will use relative IPC as the metric.
In place of coverage and accuracy, which are commonly used for prefetching schemes,
we use CWKI and Accesses per KI.
Metrics like coverage, miss rate, and MPKI don't take timing and overlap of operations into account, making the overall time that the fetch process is waiting difficult to obtain. 
CWKI takes the timing of the delivery of instruction blocks into account, and directly measures the overall impact on the fetch process. 
%\textcolor{red}{accounting for the timely delivery of instructions to the processor 
%(mentioning unlike MPKI may not be necessary because often even reviewers point out that MPKI is not a great metric with schemes such as FDIP relying on latency tolerance)} and 
Similarly, Accesses per KI directly quantifies the overhead in additional L3 accesses and traffic.

%compared to presenting a raw coverage or accuracy.
%For example, a scheme having a coverage of 95\% may bring down MPKI of 10 to 0.5 and 
%another scheme with same coverage may bring down MPKI of 100 to 5, a much larger magnitude.

\subsection {Default IPU Structures and Operation }

The default structures and sizes used by the IPU are shown in Table \ref{fig:presend-storage-overhead}.
%later in section \ref{sec:AltStrucSizes} we consider structure sizes.
An FT entry has 2B for the block address, 0.5B for the block count, 1B for the instruction count,
1.5B for the next fragment, for each of the call and return fragments,
and 1.75B for the tags/valid bit.
A DTT entry has 1.5B for the next fragment, 2 aging bits per path, and 1.25B for tags/valid bit. A BTT entry contains 3 bits to represent the block temperature.
We use three ORT tables: ORT-2, ORT-4, and ORT-16, which track 2, 4, and 16 code regions per fragment, respectively.
All tables have 8-way associativity with random replacement. MTT (multi-target table) is not used, and we limit the number of paths to two at any point of time. 
%We use 8-way associativity for the FT, DTT, and ORTs.
%This associativity is useful in avoiding conflicts when populating
%the tables. We use random replacement for all tables.
With the additional storage for the IPUS and the various queues, this totals to about 68KB (plus an additional 19KB for iTLB and BTB send) of storage with 4K FT entries,
and about 43KB (plus 10KB for iTLB and BTB send) with 2K FT entries.
Note that this storage is in (slower) secondary structures as compared to Shotgun and EP where it is in (faster) primary structures.

%iTLB enhancement accounts for about 19KB of extra storage.  (THIS ISN'T GOING TO BE NEEDED)

%Only one of the ORT tables is used per FT fragment entry, if needed, i.e., the tables are not cascaded.
%Further, when there is a need for an ORT (the OF bit is set), the Blocks field of the FT entry are used as a pointer
%to one of the ORT entries instead of holding block information for a code region.
%This allows the ORT entry to
%be accessed directly from the FT entry, without associative access.  

%The negative of this is that
%the ORTs have to hold block information for an additional region that could have been held in the FT,
%thus increasing the demand placed on the ORTs.

%For Send, we have two configurations, \textit{SendB}, which employs a 2K-entry (768B) BTT, and \textit{SendA}, which does not use a BTT.
%When not explicitly mentioned, Send refers to SendA.
We do not use a Pseudo Inclusion Bit table for the simulation results presented (for a fair comparison against other schemes).  Rather we simply probe the L1i to check for the 
presence of a block, like other prefetching schemes. 
For an L1i miss latency of 20 cycles, with a 6-wide processor, the IPU works to keep 6x20=120 instructions ahead (We evaluate sensitivity to this parameter in Section \ref{Misc_Results}). 

%\textcolor{red}{(Different keep ahead distances are briefly explored in Section \ref{Misc_Results})}  We could have this, but don't for any other metric.  What we are describing, after all, is a default configuration.  We don't mention all the variations to the default when mentioning the default.

%It does so by converting the cycle count to an approximate instruction count by
%multiplying by the fetch width. So, in our simulations, IPU keeps 140 instructions ahead for a L1 miss latency of 20 cycles. 

% Our results are unchanged for even smaller instruction lookaheads of around 60-80.

%\subsection {Result Metrics and Overview}
%\label{ResultsOverview}

%Also, talk about ITLB not needing prefetch buffer. ?)

%For speedup we will use relative IPC as the metric.
%In place of coverage and accuracy, which are commonly used for prefetching schemes,
%we use MPKI and Accesses Per KI. This is because MPKI and Accesses Per KI help see the magnitude of reduction
%better compared to presenting a raw coverage or accuracy.
%For example, a scheme having a coverage of 95\% may bring down MPKI of 10 to 0.5 and 
%another scheme with same coverage may bring down MPKI of 100 to 5, a much larger magnitude.

%Though coverages are the same, also helps to see the magnitude of reductions, which is why we resort to the usage of these metrics.

%\subsection {Results Overview}
%\label{ResultsOverview}
%
%We will 
%Here we have an overview of the results of the upcomign sections.

%\begin{table}[]
%\centering
%\input{tables/motivating_stats.tex}
%\caption{Motivating Statistics}
%\label{fig:mot-stats}
%\end{table}

We start by evaluating the redirects for FDIP, Shotgun, and PS in section \ref{Redirects}.  The redirects influence
%of the program \textcolor{green}{sequencing} determine 
how far ahead in the instruction a scheme is able to (correctly) get and the lower frequency of redirects due to high-level program map traversal is the main reason presending is able to get a better (timely and correct) stream of the instruction blocks over prefetching. We limit the number of tracks pursued by PS at any given time to two and provide an empirical evaluation in Section \ref{Redirects} demonstrating why this suffices.   After quantifying the efficacy of the high-level program map traversal, we present an evaluation of instruction presending, first presenting some operational characteristics and then evaluating the supply to different primary structures (BTB, L1i, and iTLB)

%\textcolor{red}{First, we evaluate key characteristics of the high level program sequencing in Section \ref{Redirects} which is central to the performance of the technique, before showing the evaluation for supply to different structures (iTLB, BTB and L1i).}

\begin{figure*}[h]
  \centering

  \includegraphics[width=0.6\linewidth]{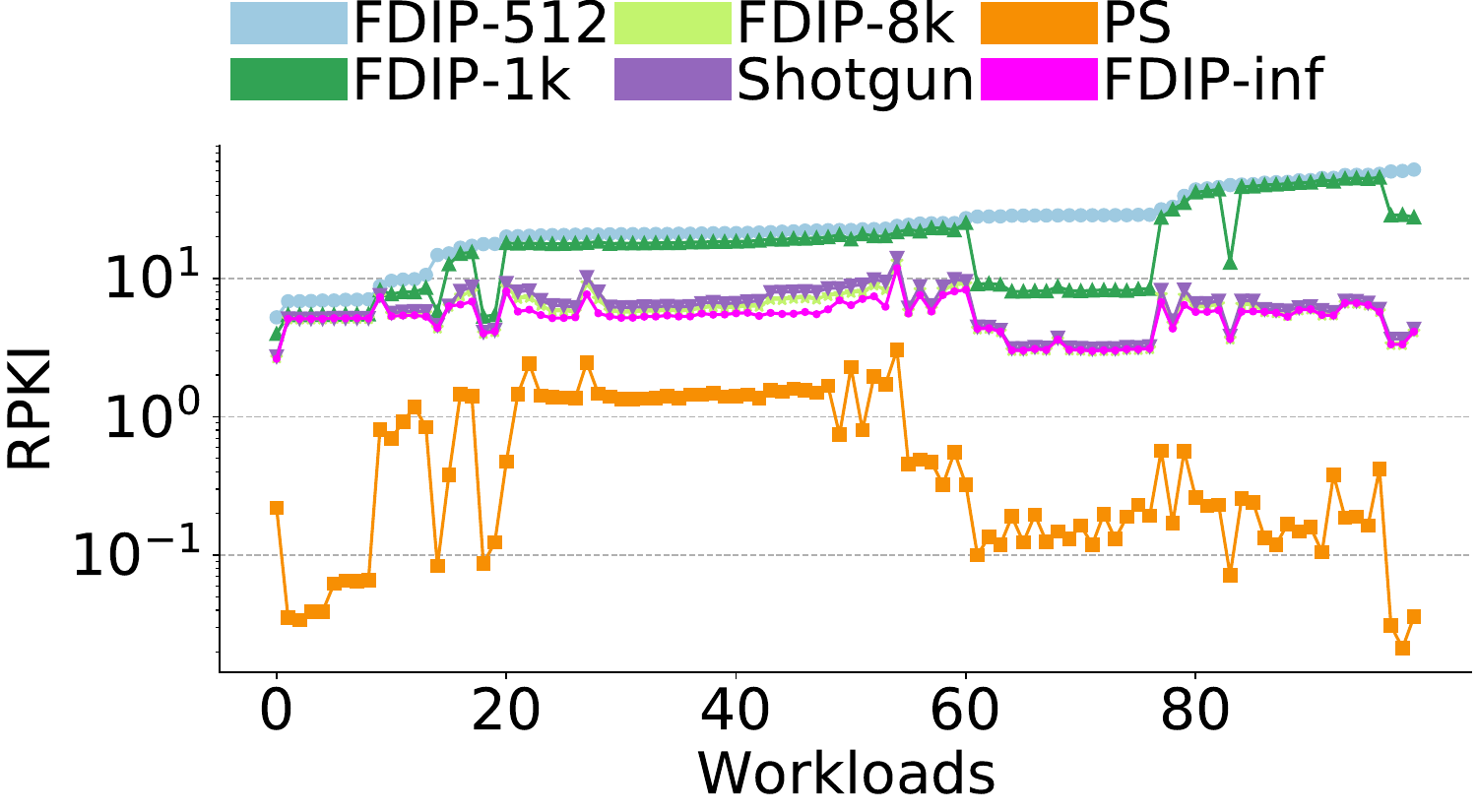}
	\caption {Redirects per Kilo Instructions (RPKI)}
  %\vspace{-20pt}
\label{fig:redirects-per-ki}
\end{figure*}

\subsection {Redirects and Select Operational Characteristics}
\label{Redirects}
 
Figure \ref{fig:redirects-per-ki} presents the \textit{redirects per KI (RPKI)} for FDIP with different sized BTBs, Shotgun, and PS. We do not show the redirects for EP, as this metric is relevant only for schemes that explicitly traverse the program control flow.
The Y axis (\textit{note the log scale}) is the RPKI, and the X axis are the 100 different benchmarks.
The first (well-known) observation is that FDIP needs large BTBs (8K) to reduce redirects.
But even with an infinite BTB, the RPKI is in the single digits, due to branch mispredictions.
For Shotgun, the RPKI is similar to the RPKI with an 8K BTB.
For PS, it is much lower, over an order of magnitude in many cases.
This demonstrates that, because of because of traversing the program map at a high level, without being perturbed by low-level control flow errors (branch mispredicts and BTB misses), PS can proceed correctly much further in discovering the instruction blocks needed, than FDIP or Shotgun.

Table \ref{fig:rpki} presents some operational characteristics of PS for the different application bins.
The first data row (RPKI) shows the RPKI.
RPKI is very low for bins I, III, IV, and V and somewhat higher for bin II.

The second data row (Br-Misp-Uniq-Succ) measures the percentage of fragments containing a branch that is mispredicted and that have a unique successor fragment. For these fragments, the high-level program map traversal (from fragment to fragment) is unaffected by the presence of mispredicted branches within the fragment.  This percentage is quite high, on average 80\% across all the bins.
The third data row (Br-Misp-Succ-InTable) measures the number of fragments enclosing a branch mispredict that have their successor fragment entered in either the FT or DTT, i.e., have one or two recorded successor fragments, as a percentage of total dynamic fragments. This percentage is about 99\% or higher across the different bins. This indicates that even in the presence of branch mispredicts in the processor, the IPU has information to operate without a redirect, by pursuing one or two tracks, for about 99\% of the dynamic fragments.

\begin{table}[]
\centering
 %\vspace{-3pt}
	\caption{PS Operational Characteristics}

\scalebox{1.0}{
\begin{tabular}{cccccc}
%\textbf{}       & \multicolumn{5}{c}{\textbf{MPKI}}                                                \\\hline
	\textbf{Metric} & \textbf{(I)} & \textbf{(II)} & \textbf{(III)} & \textbf{(IV)} & \textbf{(V)} \\\hline\hline
	RPKI  & 0.16          & 1.10           & 0.21           & 0.17           & 0.11        
	\\
 Br-Misp-Uniq-Succ        & 84.6          & 82.1          & 81.8          & 79.7        & 79.5       \\ 
  Br-Misp-Succ-InTable           & 98.7          & 98.7          & 98.9          & 99.6        & 99.7       \\ 
 
	Br/fragment          & 5.4          & 5.2          & 4.6          & 4.2          & 3.6        \\ 
	
	%Instr/fragment           & 30.1          & 27.8          & 26.1          & 24.3          & 19.5       \\ 
	Update           & 0.17          & 0.44          & 0.14          & 0.11          & 0.07       \\\hline 
	ORT-2           & 21.21         & 21.52          & 13.74          & 12.68          & 10.56         \\
	ORT-4          & 2.53          & 1.63           & 0.60           & 0.41           & 0.11          \\
	ORT-16         & 0.05          & 0.04           & 0.07           & 0.07           & 0.01          \\
	DTT             &6.11          & 9.49           & 6.69           & 5.69           & 4.06         
\end{tabular}
}

\label{fig:rpki}

\end{table}

The fourth data row (Br/fragment) shows the average branches per fragment.
There are on average 3.5-5.5 branches per fragment, which minimally impact RPKI for PS.
FDIP must correctly get past those many branches in a fragment (correct direction prediction and BTB hit)
to get to the (correct) instruction blocks that PS is able to get to.

%The fifth data row (Instr/fragment ) has the average number of instructions per fragment.
%If the processor sends the identifier of every fragment it fetches to the IPU,
%this number is indicative of the communication rate.
%For bin II, the processor will have to send, on average, one fragment ID per 27.8 instructions,
%or once every 10 cycles if the processor is achieving a processing speed of 2.78 IPC.

The fifth data row (Update) is the percentage of fragments processed that are involved in the creation or update
of an FT entry.  For example, for bin II, 0.44\% of the fragments update the FT; 99.56\% do not, i.e., once an FT entry is created/updated, it is used about 225 times.
For bin V, this number is over 1400.
The low FT update rate (Update) indicates a low rate of communication from the processor to create and maintain the FT.

%Additionally, a low RPKI can be achieved with a low fragment identifier communication (Instructions/fragment) rate.
%Collectively these numbers suggest that the IPU can proceed productively in a mostly decoupled fashion from the processor.

The last data row (DTT) quantifies the percentage of FT accesses which also involves a DTT lookup.
Observe that this percentage is quite small (most FT accesses have a unique next fragment).
The remaining data rows present the percentage of FT accesses which involve a look-up to ORT-2, ORT-4, and ORT-16, respectively.
%(different overflow region tables, holding onto discontiguous regions of a fragment), respectively.
Observe that these percentages are quite small for ORT-4/ORT-16.
They are a little higher for bins I and II because benchmarks in these bins have more branches
and thus more local control flow within a fragment.

%Table \ref {fig:itlb-acc} presents the average FPKI for the different application bins for FDIP-32, Send-32, FDIP-64, Send-64, which are the Base and Send configurations,
%with 32 and 64 entry iTLBs, respectively.
%In Bins I and II, which have low base MPKI, there is about 40\% additional traffic (FPKI) with Send; less so for the higher base MPKI benchmarks in Bins III, IV, and V.

%This section we will look at ILTBs for a fixed FDIP configuration with 8K BTB entries and 64 entry L1iTLB.
%
%
%Here we will have graphs for ITLB speedups over FDIP-8K, 64-oTLB with larger (128, 256) TLBs for FDIP,
%and also with smaller TLBs (16, 32, 64, infinite) with Send.

%Then also have the MPKI for the different TLBs.

%And how much additional traffic (fills) for 16, 32, 64 TLBs with Send over base 64 TLB.  This can be done as figures as we have it currently.

\begin{figure}[h]
  \includegraphics[width=1.0\linewidth]{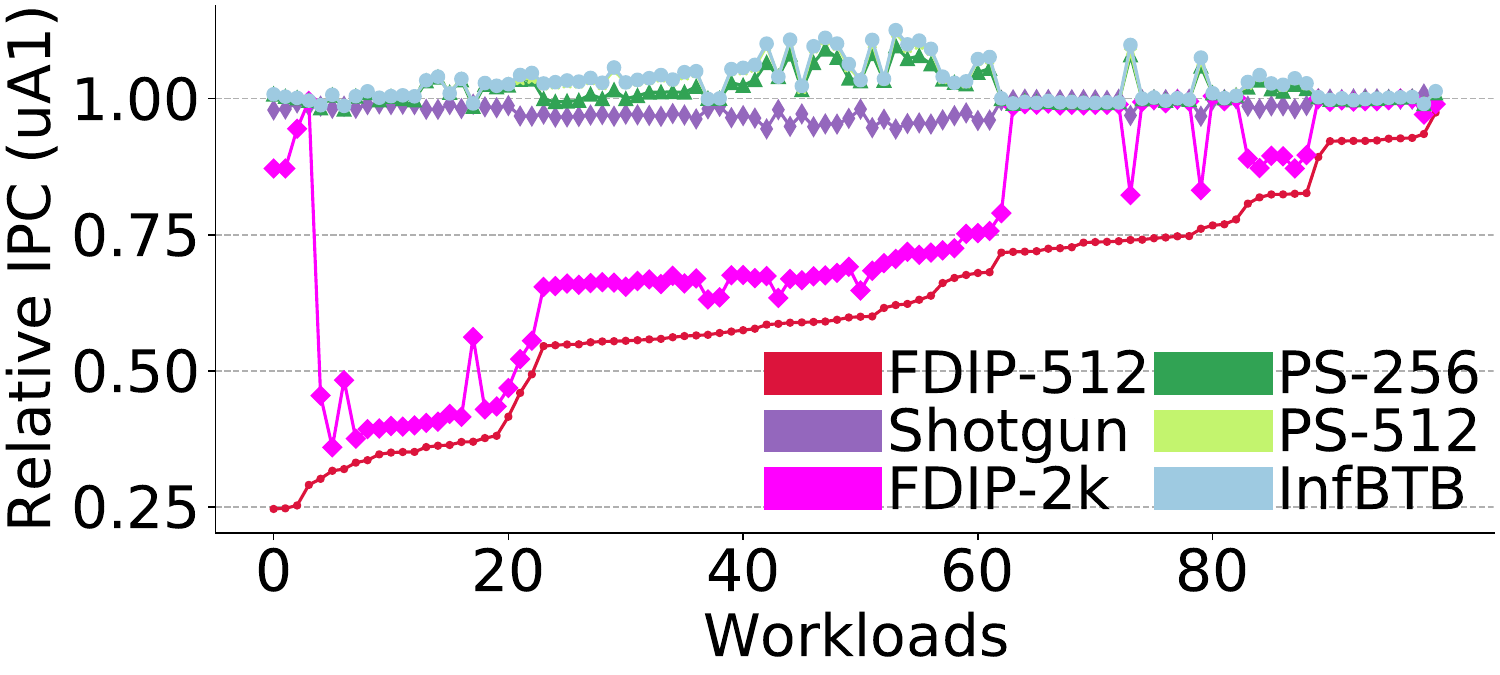}
	\caption {Relative IPC BTB Presending}
  \label{fig:send-btb-entries}
  %\vspace{-20pt}
\end{figure}

%\begin{table}[]
%\centering
%\input{tables/btb_mpki_var.tex}
%\caption{BTB MPKI}
%\label{fig:btb-mpki}
%\end{table}
%
%
%\begin{table}[]
%\centering
%\input{tables/btb_fills_var.tex}
%\caption{BTB FPKI}
%\label{fig:btb-fills}
%\end{table}

\subsection {Branch Target Buffers}
\label{BTB_Results}

We now consider PS for BTB entries alone.
Figure \ref{fig:send-btb-entries} presents the IPC of several configurations, relative FDIP  with an 8K BTB.
The graph presents the relative IPC for FDIP with 512, 2K, and $\infty$ BTB entries, Shotgun, and PS with 256 (PS-256) and 512 (PS-512) BTB entries.
FDIP alone with smaller BTB sizes suffers significant performance loss, as is already known.
The Shotgun configuration simulated is close to the 8K BTB base, slightly lower for some benchmarks.
PS, on the other hand, achieves better performance with only 256 BTB entries than FDIP with 8K BTB entries.
With 512 BTB entries, PS is close to FDIP with an infinite BTB.
Conventional wisdom is that large primary BTBs are needed for large code footprint workloads.
%Very large BTBs have significant implementation issues, especially at high clock speeds.
With PS, even small primary BTBs suffice, as PS can move the necessary BTB entries from the secondary BTB to the primary BTB in a timely fashion. For all upcoming evaluations for PS, we use a 512-entry L1 BTB, and a 16K entry, 8cycle L2 BTB (results with 8K entry L2 BTB are very similar).

%Table \ref{fig:btb-mpki} and Table \ref{fig:btb-fills}
%Here we will have the main BTB effectiveness results for base FDIP and +Send.
%
%Can have the IPC relative to FDIP-8K for FDIP 512, 1K, etc. and also with Send (256, 512) as a graph (as is currently).
%
%Then have tables for different characteristics, with the columns for different MPKIs.
%Here we could have MPKI, IPC, Fills for different BTB sizes for FDIP and also Send.

%\begin{table}[]
%\centering
%\input{tables/icache_accuracy.tex}
%\caption{Icache FPKI}
%\label{fig:icache-acc}
%\end{table}
\begin{figure*}[h]
  \centering
  \includegraphics[width=0.6\linewidth]{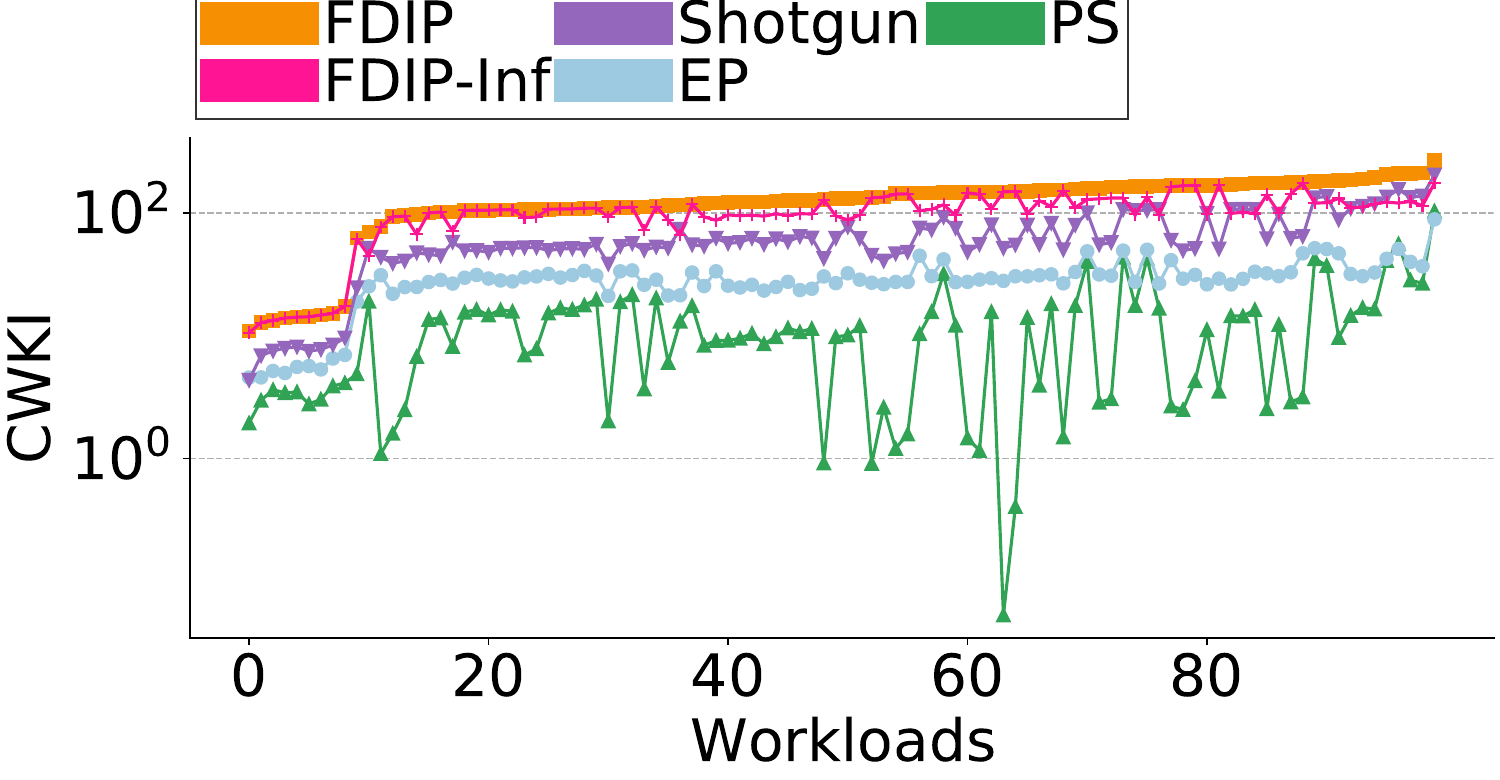}
	\caption {Cycles Waiting per KI (CWKI)}
  \label{fig:send-blks-comparisons-mpki-ua1}
\end{figure*}

\begin{figure}[h]
  \includegraphics[width=1.0\linewidth]{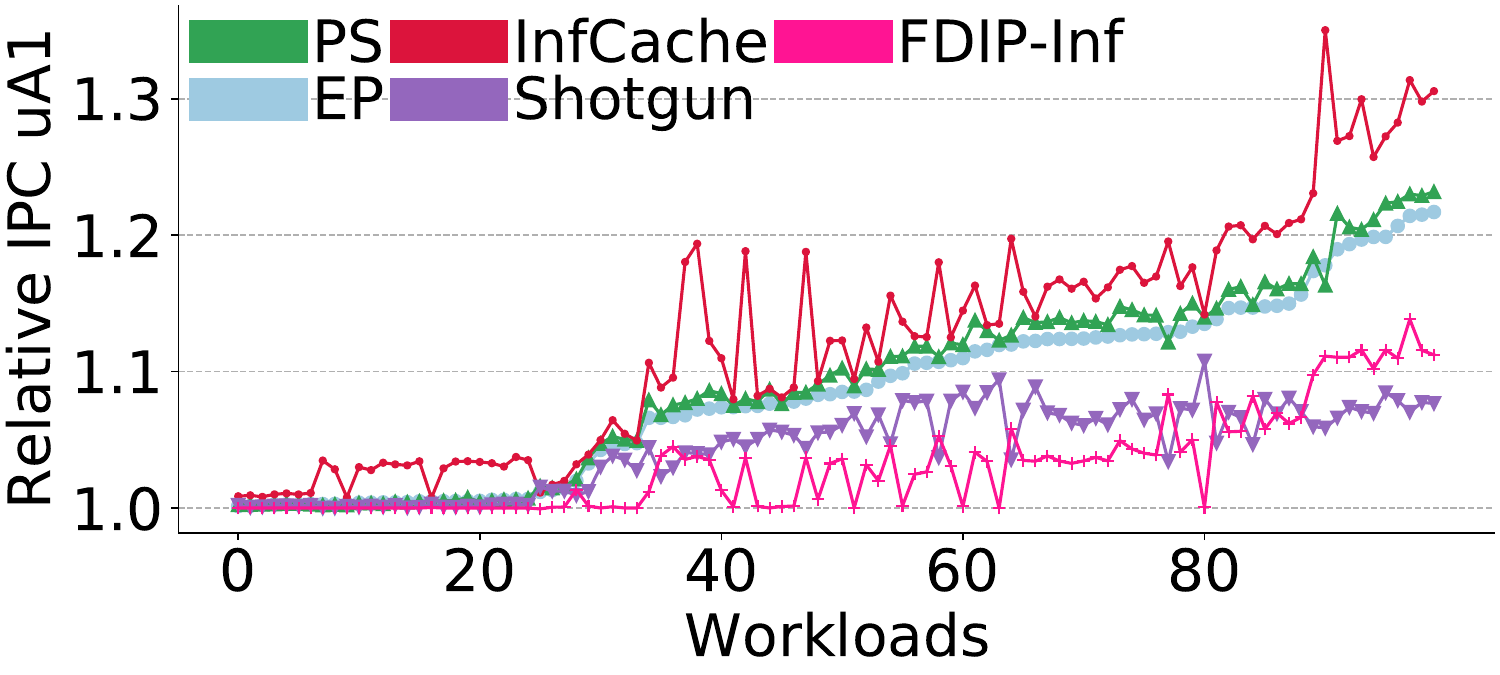}
	\caption {Relative IPC uA1}
  \label{fig:send-blks-comparisons-ipc-ua1}
\end{figure}

\begin{figure}[h]
  \includegraphics[width=1.0\linewidth]{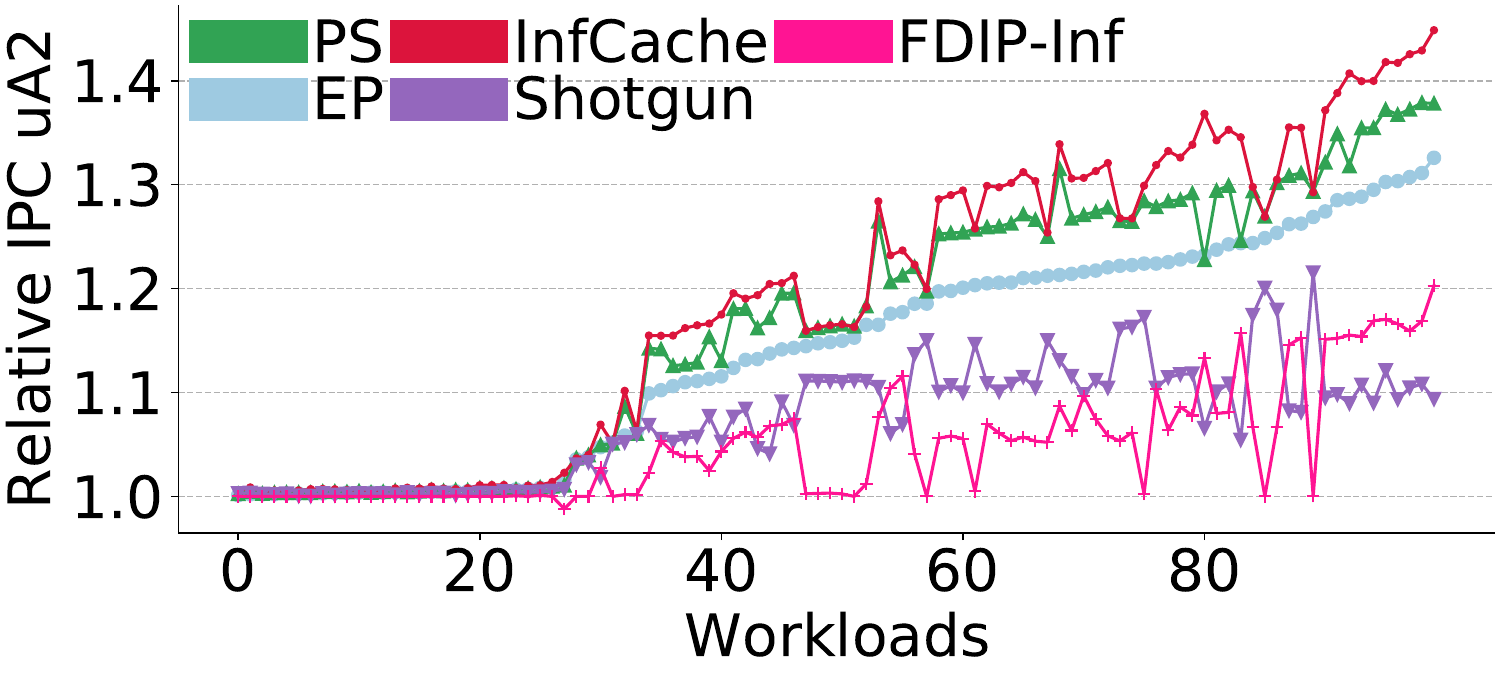}
	\caption {Relative IPC uA2}
  \label{fig:send-blks-comparisons-ipc-ua2}
 %\vspace{-10pt}
\end{figure}
\subsection {L1i Cache }
\label{L1Cache-CoreResults}

We now evaluate PS for the L1i cache for both uA1 and uA2 as the aggressive back end of uA2 demands greater L1i efficiency.
For FDIP we use a primary 8K BTB (FDIP) and an infinite BTB (FDIP-Inf).
For Entangling Prefetch (EP), we also use an 8K primary BTB.
EP's effectiveness for L1i misses isn't impacted much by the primary BTB size; indeed the L1i MPKI
for EP with a smaller BTB (e.g., 512 entries), not shown, is also very good.
But the overall IPC gets impacted because the smaller BTB, without additional management,
impacts the fetching (and thus the IPC), significantly.
The (aggressive) Shotgun configuration is as described in section \ref{SimDetails}.
We also consider an infinite L1i cache (InfCache). Although we use a 512-entry primary BTB for PS, our results shown in Figures \ref{fig:send-blks-comparisons-mpki-ua1}, \ref{fig:send-blks-comparisons-ipc-ua1}, and \ref{fig:send-blks-comparisons-ipc-ua2} are mostly similar with an 8K-entry primary BTB without BTB management. We primarily present results with a smaller BTB to demonstrate that PS enables high performance with a smaller primary BTB and effective BTB management.

%We do not use a shadow cache.  Rather we simply probe the L1i to check for the 
%presence of a block, similar to other prefetching schemes. When not explicitly mentioned, Send refers to SendA.

%Here we also compare with \textit{Entangling Prefetch (EP)}, \textcolor{red}{whose storage overhead is 77.44KB}. For Send, we do the evaluation using a \textbf{512 entry BTB}, whereas for EP we do the evaluation using \textbf{8K BTB}. We could use a smaller BTB for EP, but we use 8K BTB for a fair comparison.  

Figure \ref{fig:send-blks-comparisons-mpki-ua1} presents the CWKI (log scale) for the different configurations.
FDIP, with an 8K BTB, or with an infinite BTB, and Shotgun have a similar CWKI.
EP achieves a much lower CWKI, but PS is even lower, by an order of magnitude for some of the benchmarks.
There is still some improvement with an infinite cache.
CWKI trends and magnitudes for uA2, not shown, are similar.

%\textcolor{red}{Send which explores only the previous successor fragment (Send-Prev) is also studied here. Send-Prev helps bring down the misses a lot over EP (for some benchmarks coming very close to SendA and SendB), leaving a small room for improvement which Send can cover by exploring the additional track.}  

%We also present the MPKI with uA2 for FDIP, Send, EP, and an infinite cache, and observe 
%a similar trend, though the MPKIs have increased a little for Send and EP with the aggressive backend.

These lower CWKIs translate into increased performance, as shown in
Figure \ref{fig:send-blks-comparisons-ipc-ua1} for uA1 and
Figure \ref{fig:send-blks-comparisons-ipc-ua2} for uA2.
In these figures, IPC is shown relative to the base FDIP (with an 8K BTB) values.
A reason that the aggressive Shotgun can 
achieve a better IPC than FDIP, even with an infinite BTB, is
because we allow it to prefetch blocks from multiple code regions in a fragment that spans multiple UBTB entries in one step, whereas FDIP would take multiple steps to achieve the same.
EP does quite well, performing better than FDIP and Shotgun.
For uA1, PS perform only slightly better than EP, even though the CWKI is much lower.
For uA2, because of the higher demands for front-end efficiency, the performance difference is greater,
especially for applications where PS has much lower CWKI than EP.
The performance improvement over EP in several benchmarks is 2-6\%. Although the performance improvement over EP is modest, PS achieves comparable performance with a smaller BTB and effective BTB management.

\subsection {L3 Traffic and L1i Pollution Overheads}
\label{L3Traffic}

\begin{figure}[h]
  \includegraphics[width=1.0\linewidth]{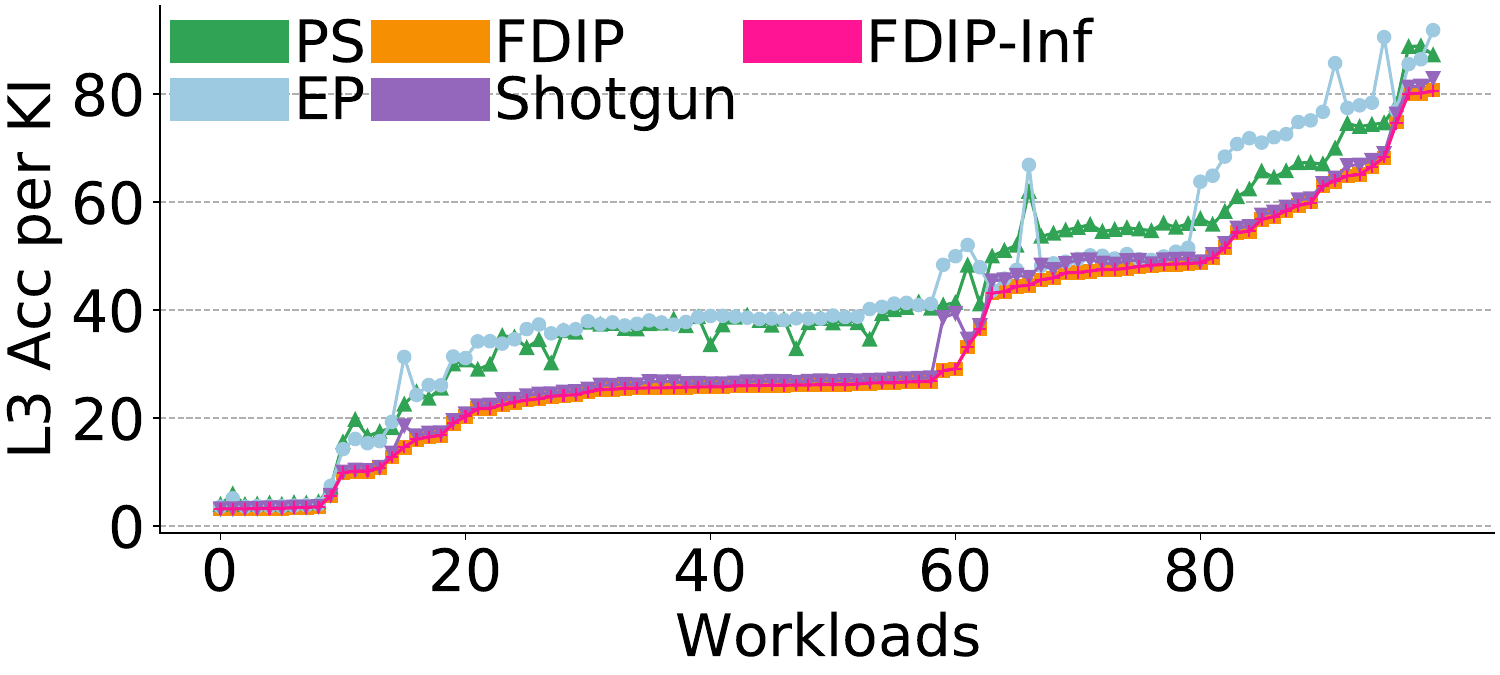}
	\caption {Traffic Increase for uA1}
  \label{fig:traffic-increase}
 %\vspace{-10pt}
\end{figure}

Figure \ref{fig:traffic-increase} quantifies the traffic as L3 accesses per KI introduced by PS, EP, Shotgun, and the baseline FDIP,
for which the L1i CWKIs and IPCs were quantified in section \ref{L1Cache-CoreResults}.
Here the results for FDIP are somewhat optimistic as we do a trace-driven simulation, and wrong-path prefetches are not counted in the traffic.
Shotgun's traffic is like FDIP, and slightly worse in some cases, for the reason mentioned above.
The traffic for PS is higher than for FDIP (and Shotgun) in almost all cases though the difference is relatively smaller in cases where the misses were higher.
Note that the traffic for PS is on par with that of EP,
which is more accurate than other prefetching schemes \cite{ros2021cost}.
As has been noted by \cite{gober2020temporal},  for most of these traces, many L1i blocks are dead.  Though PS and EP move additional (sometimes dead) blocks into the L1i, these additional blocks mostly replace dead blocks, and thus they can continue to maintain a low MPKI and CWKI despite the additional block movement.

\subsection {IPU Keep Ahead Distance }
\label{Misc_Results}

Table \ref{fig:cache-size-var-uarch2} presents the average CWKI as we vary the distance, measured in number of instructions, the IPU keeps ahead of the processor for the 5 different application bins.
%Row FDIP-MPKI presents the L1i MPKI for FDIP and rows S32-40-MPKI, S32-80-MPKI, S32-120-MPKI, and S32-160-MPKI present the L1i MPKIs with 
Rows PS-40, PS-80, PS-120, and PS-160 present the CWKIs with PS trying to keep 40, 80, 120, and 160 instructions, 
respectively, ahead of the processor.  
In program phases where 6 instructions are being fetched every cycle, keeping 120 (6x20) instructions ahead is needed to tolerate a 20-cycle latency. 
However, such a fetch rate is rarely sustained throughout the execution. 
Keeping only 40 instructions ahead achieves a low CWKI, much better than the base FDIP (not shown), however there is more room for improvement.  Keeping 80 instructions ahead reduces CWKI even more.  Beyond 120 instructions, at 160 instructions, the CWKI is nearly the same.
%with a very slight increase for bin II, which has more fragments with two targets (DTT accesses, see Table \ref{fig:rpki}), due to more wasteful blocks sent from two paths. 

\begin{table}[]
\centering
	
\caption{CWKI for Varying PS Keep Ahead Distance}

\scalebox{1.0}{
\begin{tabular}{cccccc}
   %  & \multicolumn{5}{c}{\textbf{uA1}}                                              \\\hline
     & \textbf{(I)} & \textbf{(II)} & \textbf{(III)} & \textbf{(IV)} & \textbf{(V)} \\\hline\hline
PS-40  & 9 & 15 & 13 & 10 & 3 \\
PS-80  & 7 & 14 & 12 & 7  & 2 \\
PS-120 & 6 & 13 & 11 & 7  & 1 \\
PS-160 & 6 & 13 & 10 & 6  & 1 \\\hline

%S32L30   & 2.16 & 2.55 & 1.89 & 1.91 & 2.35 \\
%S32L40   & 2.13 & 2.49 & 1.86 & 1.90 & 2.34 \\\hline
\end{tabular}
}
%\begin{tabular}{cccccc}
%\textbf{Config} & \textbf{(I)} & \textbf{(II)} & \textbf{(III)} & \textbf{(IV)} & \textbf{(V)} \\\hline\hline
%S8              & 2.18         & 2.30          & 1.95           & \textbf{2.02}          & \textbf{2.48}         \\
%S16             & 2.24         & 2.41          & 1.97           & \textbf{2.03}          & \textbf{2.50}         \\
%S32             & 2.25         & 2.49          & 1.98           & \textbf{2.03}          & \textbf{2.50}         \\
%	EP32 & 2.22 & 2.37 & 1.88 & \textbf{1.97} & \textbf{2.35} \\
%S32L30          & 2.22         & 2.45          & 1.94           & 2.02          & 2.49         \\
%S32L40          & 2.18         & 2.39          & 1.90           & 2.00          & 2.46         \\
%Inf             & 2.27         & 2.56          & 2.00           & \textbf{2.03}          & \textbf{2.51}        
%\end{tabular}
\label{fig:cache-size-var-uarch2}

% \vspace{10pt}
\end{table}

\begin{figure}[h]
  \includegraphics[width=1.0\linewidth]{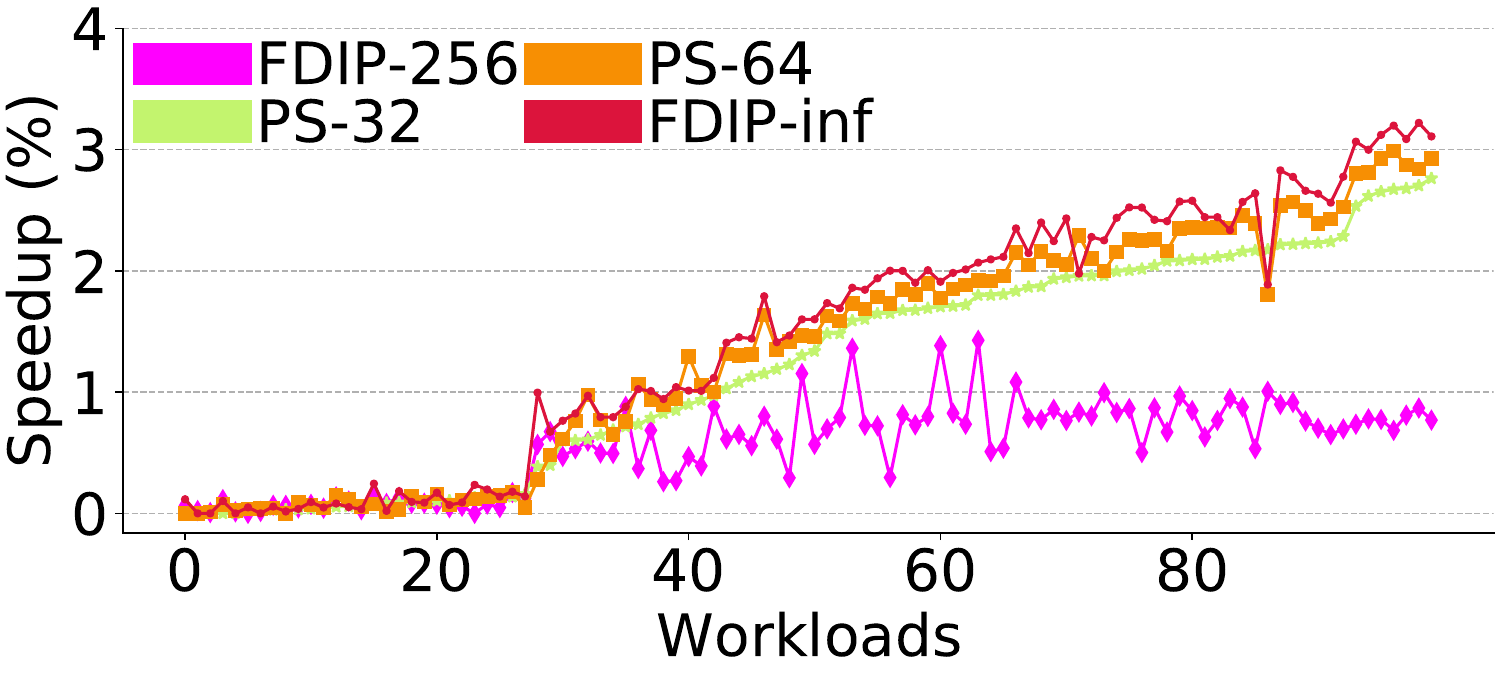}
	\caption {iTLB Speedups}
  \label{fig:send-itlb-entries-ipc}

  %\vspace{-25pt}
\end{figure}

\subsection {Instruction TLBs}
\label{ITLB_Results}

Finally, we consider PS for iTLBs alone, for uA1, using PS to move from a 2K entry secondary (L2) TLB to a primary iTLB.
%Figure \ref{fig:send-itlb-entries-mpki} 
%presents the iTLB MPKI (log scale) for FDIP (8K entry BTB) for 64-, 256- and infinite-entry iTLBs (FDIP-64, 256, inf),
%without additional prefetching,
%and for Send with 32- and 64-entry iTLBs.
%Note the significantly smaller iTLB MPKI with Send, two orders of magnitude in several cases for a 64-entry iTLB.
Figure \ref{fig:send-itlb-entries-ipc},
which presents the speedup over a base case of FDIP with a 64-entry iTLB for: (i) FDIP with a 256-entry iTLB (FDIP-256),
(ii) PS with a 32-entry iTLB (PS-32), (iii) PS with a 64-entry iTLB (PS-64), and (iv) FDIP with an infinite iTLB (FDIP-Inf).
With PS, the primary iTLB misses are reduced significantly (not shown), resulting in the speedups shown in the figure.
PS with a 32-entry iTLB does better than even a normal 256-entry iTLB configuration, and PS with a 64-entry iTLB is close to
a normal infinite iTLB configuration.

\section{Concluding Remarks}
\label{sec:Conclusions}

This paper \textit{instruction presending}, where code cache blocks, iTLB and BTB
entries, are proactively moved from secondary structures to primary structures.
%by traversing a high-level program map.
Working with tables holding a high-level program map, 
%containing information about fragments of a program, 
and without branch predictors and BTBs, the presending hardware operates autonomously to send
the requisite information (iTLB and BTB entries, and instruction cache blocks) to the processor, in a timely manner.

Presending is very effective, especially for large code footprint applications,
achieving an order of magnitude lower MPKI than state-of-the-art instruction prefetching techniques in many cases, enabling the processor to achieve equivalent or higher performance with much smaller-sized primary BTBs. Effective instruction presending also opens up the possibility of sending additional information about the instruction stream (possibly obtained via preprocessing), that can be used to further improve instruction processing.

%\textcolor{red}{Other applications of the high-level sequencing mechanism are also left to future work.}

%This initial paper only studied a small part of the overall design space for instruction presending.
%Many other aspects need to be studied, especially to improve the quality of
%the blocks sent for applications when there are multiple possible next fragments.
%The choice of how many (and which) paths the IPU should pursue is an important area for further investigation.
%Alternate means for creating the shadow program software, is another interesting research avenue.

%Say something about FT can be shared among different cores running a common workload with code kept in L3 and not replicated, leaving L2 to data..

\bibliographystyle{plain}
\bibliography{refs}

@inproceedings{bonanno2013two,
  title={Two level bulk preload branch prediction},
  author={Bonanno, James and Collura, Adam and Lipetz, Daniel and Mayer, Ulrich and Prasky, Brian and Saporito, Anthony},
  booktitle={2013 IEEE 19th International Symposium on High Performance Computer Architecture (HPCA)},
  pages={71--82},
  year={2013},
  organization={IEEE}
}

@inproceedings{asheim2023storage,
  title={A Storage-Effective BTB Organization for Servers},
  author={Asheim, Truls and Grot, Boris and Kumar, Rakesh},
  booktitle={2023 IEEE International Symposium on High-Performance Computer Architecture (HPCA)},
  pages={1153--1167},
  year={2023},
  organization={IEEE}
}

@article{gober2022championship,
  title={The championship simulator: Architectural simulation for education and competition},
  author={Gober, Nathan and Chacon, Gino and Wang, Lei and Gratz, Paul V and Jimenez, Daniel A and Teran, Elvira and Pugsley, Seth and Kim, Jinchun},
  journal={arXiv preprint arXiv:2210.14324},
  year={2022}
}

@inproceedings{wang2023acic,
  title={Acic: Admission-controlled instruction cache},
  author={Wang, Yunjin and Chang, Chia-Hao and Sivasubramaniam, Anand and Soundararajan, Niranjan},
  booktitle={2023 IEEE International Symposium on High-Performance Computer Architecture (HPCA)},
  pages={165--178},
  year={2023},
  organization={IEEE}
}

@inproceedings{brunner2024weeding,
  title={Weeding out Front-End Stalls with Uneven Block Size Instruction Cache},
  author={Brunner, Roman and Kumar, Rakesh},
  booktitle={2024 57th IEEE/ACM International Symposium on Microarchitecture (MICRO)},
  pages={1382--1396},
  year={2024},
  organization={IEEE}
}

@inproceedings{liu2024avm,
  title={AVM-BTB: Adaptive and Virtualized Multi-level Branch Target Buffer},
  author={Liu, Yunzhe and Li, Xinyu and Zhang, Tingting and Liu, Tianyi and Guo, Qi and Zhang, Fuxin and Wang, Jian},
  booktitle={2024 ACM/IEEE 51st Annual International Symposium on Computer Architecture (ISCA)},
  pages={17--31},
  year={2024},
  organization={IEEE}
}

@article{burcea2008predictor,
  title={Predictor virtualization},
  author={Burcea, Ioana and Somogyi, Stephen and Moshovos, Andreas and Falsafi, Babak},
  journal={ACM SIGOPS Operating Systems Review},
  volume={42},
  number={2},
  pages={157--167},
  year={2008},
  publisher={ACM New York, NY, USA}
}

@inproceedings{pnevmatikatos1993control,
  title={Control flow prediction for dynamic ILP processors},
  author={Pnevmatikatos, Dionisios N and Franklin, Manoj and Sohi, Gurindar S},
  booktitle={Proceedings of the 26th Annual International Symposium on Microarchitecture},
  pages={153--163},
  year={1993},
  organization={IEEE}
}

@inproceedings{adiga2020ibm,
  title={The ibm z15 high frequency mainframe branch predictor industrial product},
  author={Adiga, Narasimha and Bonanno, James and Collura, Adam and Heizmann, Matthias and Prasky, Brian R and Saporito, Anthony},
  booktitle={2020 ACM/IEEE 47th Annual International Symposium on Computer Architecture (ISCA)},
  pages={27--39},
  year={2020},
  organization={IEEE}
}

@inproceedings{khan2021twig,
  title={Twig: Profile-guided btb prefetching for data center applications},
  author={Khan, Tanvir Ahmed and Brown, Nathan and Sriraman, Akshitha and Soundararajan, Niranjan K and Kumar, Rakesh and Devietti, Joseph and Subramoney, Sreenivas and Pokam, Gilles A and Litz, Heiner and Kasikci, Baris},
  booktitle={MICRO-54: 54th Annual IEEE/ACM International Symposium on Microarchitecture},
  pages={816--829},
  year={2021}
}

@inproceedings{song2022thermometer,
  title={Thermometer: profile-guided btb replacement for data center applications},
  author={Song, Shixin and Khan, Tanvir Ahmed and Shahri, Sara Mahdizadeh and Sriraman, Akshitha and Soundararajan, Niranjan K and Subramoney, Sreenivas and Jim{\'e}nez, Daniel A and Litz, Heiner and Kasikci, Baris},
  booktitle={Proceedings of the 49th Annual International Symposium on Computer Architecture},
  pages={742--756},
  year={2022}
}

@inproceedings{chaconsw,
  title={A Characterization of the Effects of Software Instruction Prefetching on an Aggressive Front-end},
  author={Chacon, Gino and Gober, Nathan and Nathella, Krishna and Gratz, Paul and Jim{\'e}nez, Daniel},
  booktitle={2023 IEEE International Symposium on Performance Analysis of Systems and Software (ISPASS)},
  pages={24--34},
  year={2023},
  organization={IEEE}
}

@inproceedings{kanev2015profiling,
  title={Profiling a warehouse-scale computer},
  author={Kanev, Svilen and Darago, Juan Pablo and Hazelwood, Kim and Ranganathan, Parthasarathy and Moseley, Tipp and Wei, Gu-Yeon and Brooks, David},
  booktitle={Proceedings of the 42nd Annual International Symposium on Computer Architecture},
  pages={158--169},
  year={2015}
}

@inproceedings{ayers2019asmdb,
  title={Asmdb: understanding and mitigating front-end stalls in warehouse-scale computers},
  author={Ayers, Grant and Nagendra, Nayana Prasad and August, David I and Cho, Hyoun Kyu and Kanev, Svilen and Kozyrakis, Christos and Krishnamurthy, Trivikram and Litz, Heiner and Moseley, Tipp and Ranganathan, Parthasarathy},
  booktitle={Proceedings of the 46th International Symposium on Computer Architecture},
  pages={462--473},
  year={2019}
}

@article{kumar2018blasting,
  title={Blasting through the front-end bottleneck with shotgun},
  author={Kumar, Rakesh and Grot, Boris and Nagarajan, Vijay},
  journal={ACM SIGPLAN Notices},
  volume={53},
  number={2},
  pages={30--42},
  year={2018},
  publisher={ACM New York, NY, USA}
}

@article{gober2020temporal,
  title={Temporal ancestry prefetcher},
  author={Gober, Nathan and Chacon, Gino and Jim{\'e}nez, DA and Gratz, P},
  journal={The 1st Instruction Prefetching Championship (IPC1)},
  year={2020}
}

@inproceedings{ailamaki1999dbmss,
  title={DBMSs on a modern processor: Where does time go?},
  author={Ailamaki, Anastassia and DeWitt, David J and Hill, Mark D and Wood, David A},
  booktitle={VLDB'99, Proceedings of 25th International Conference on Very Large Data Bases, September 7-10, 1999, Edinburgh, Scotland, UK},
  number={CONF},
  pages={266--277},
  year={1999}
}

@inproceedings{cao1999detailed,
  title={Detailed characterization of a quad Pentium Pro server running TPC-D},
  author={Cao, Qiang and Trancoso, Pedro and Larriba-Pey, J-L and Torrellas, Josep and Knighten, Robert and Won, Youjip},
  booktitle={Proceedings 1999 IEEE International Conference on Computer Design: VLSI in Computers and Processors (Cat. No. 99CB37040)},
  pages={108--115},
  year={1999},
  organization={IEEE}
}

@inproceedings{keeton1998performance,
  title={Performance characterization of a quad Pentium Pro SMP using OLTP workloads},
  author={Keeton, Kimberly and Patterson, David A and He, Yong Qiang and Raphael, Roger C and Baker, Walter E},
  booktitle={Proceedings of the 25th annual international symposium on Computer architecture},
  pages={15--26},
  year={1998}
}

@inproceedings{spracklen2005effective,
  title={Effective instruction prefetching in chip multiprocessors for modern commercial applications},
  author={Spracklen, Lawrence and Chou, Yuan and Abraham, Santosh G},
  booktitle={11th International Symposium on High-Performance Computer Architecture},
  pages={225--236},
  year={2005},
  organization={IEEE}
}

@inproceedings{zhang2002execution,
  title={Execution history guided instruction prefetching},
  author={Zhang, Yi and Haga, Steve and Barua, Rajeev},
  booktitle={Proceedings of the 16th international conference on Supercomputing},
  pages={199--208},
  year={2002}
}

@inproceedings{ayers2018memory,
  title={Memory hierarchy for web search},
  author={Ayers, Grant and Ahn, Jung Ho and Kozyrakis, Christos and Ranganathan, Parthasarathy},
  booktitle={2018 IEEE International Symposium on High Performance Computer Architecture (HPCA)},
  pages={643--656},
  year={2018},
  organization={IEEE}
}

@inproceedings{kolli2013rdip,
  title={RDIP: Return-address-stack directed instruction prefetching},
  author={Kolli, Aasheesh and Saidi, Ali and Wenisch, Thomas F},
  booktitle={2013 46th Annual IEEE/ACM International Symposium on Microarchitecture (MICRO)},
  pages={260--271},
  year={2013},
  organization={IEEE}
}

@inproceedings{kumar2017boomerang,
  title={Boomerang: A metadata-free architecture for control flow delivery},
  author={Kumar, Rakesh and Huang, Cheng-Chieh and Grot, Boris and Nagarajan, Vijay},
  booktitle={2017 IEEE International Symposium on High Performance Computer Architecture (HPCA)},
  pages={493--504},
  year={2017},
  organization={IEEE}
}

@inproceedings{sriraman2019softsku,
  title={Softsku: Optimizing server architectures for microservice diversity@ scale},
  author={Sriraman, Akshitha and Dhanotia, Abhishek and Wenisch, Thomas F},
  booktitle={Proceedings of the 46th International Symposium on Computer Architecture},
  pages={513--526},
  year={2019}
}

@inproceedings{sohi1995multiscalar,
  title={Multiscalar processors},
  author={Sohi, Gurindar S and Breach, Scott E and Vijaykumar, TN},
  booktitle={Proceedings of the 22nd annual international symposium on Computer architecture},
  pages={414--425},
  year={1995}
}

@article{annavaram2003call,
  title={Call graph prefetching for database applications},
  author={Annavaram, Murali and Patel, Jignesh M and Davidson, Edward S},
  journal={ACM Transactions on Computer Systems (TOCS)},
  volume={21},
  number={4},
  pages={412--444},
  year={2003},
  publisher={ACM New York, NY, USA}
}

@inproceedings{ferdman2008temporal,
  title={Temporal instruction fetch streaming},
  author={Ferdman, Michael and Wenisch, Thomas F and Ailamaki, Anastasia and Falsafi, Babak and Moshovos, Andreas},
  booktitle={2008 41st IEEE/ACM International Symposium on Microarchitecture},
  pages={1--10},
  year={2008},
  organization={IEEE}
}

@inproceedings{kaynak2015confluence,
  title={Confluence: unified instruction supply for scale-out servers},
  author={Kaynak, Cansu and Grot, Boris and Falsafi, Babak},
  booktitle={Proceedings of the 48th International Symposium on Microarchitecture},
  pages={166--177},
  year={2015}
}

@article{ansari2021mana,
  title={Mana: Microarchitecting an instruction prefetcher},
  author={Ansari, Ali and Golshan, Fatemeh and Lotfi-Kamran, Pejman and Sarbazi-Azad, Hamid},
  journal={arXiv preprint arXiv:2102.01764},
  year={2021}
}

@article{chang1997target,
  title={Target prediction for indirect jumps},
  author={Chang, Po-Yung and Hao, Eric and Patt, Yale N},
  journal={ACM SIGARCH Computer Architecture News},
  volume={25},
  number={2},
  pages={274--283},
  year={1997},
  publisher={ACM New York, NY, USA}
}

@inproceedings{reinman1999fetch,
  title={Fetch directed instruction prefetching},
  author={Reinman, Glenn and Calder, Brad and Austin, Todd},
  booktitle={MICRO-32. Proceedings of the 32nd Annual ACM/IEEE International Symposium on Microarchitecture},
  pages={16--27},
  year={1999},
  organization={IEEE}
}

@inproceedings{srinivasan2001branch,
  title={Branch history guided instruction prefetching},
  author={Srinivasan, Viji and Davidson, Edward S and Tyson, Gary S and Charney, Mark J and Puzak, Thomas R},
  booktitle={Proceedings HPCA Seventh International Symposium on High-Performance Computer Architecture},
  pages={291--300},
  year={2001},
  organization={IEEE}
}

@inproceedings{ishii2021re,
  title={Re-establishing fetch-directed instruction prefetching: An industry perspective},
  author={Ishii, Yasuo and Lee, Jaekyu and Nathella, Krishnendra and Sunwoo, Dam},
  booktitle={2021 IEEE International Symposium on Performance Analysis of Systems and Software (ISPASS)},
  pages={172--182},
  year={2021},
  organization={IEEE}
}

@inproceedings{ferdman2011proactive,
  title={Proactive instruction fetch},
  author={Ferdman, Michael and Kaynak, Cansu and Falsafi, Babak},
  booktitle={Proceedings of the 44th Annual IEEE/ACM International Symposium on Microarchitecture},
  pages={152--162},
  year={2011}
}

@inproceedings{vavouliotis2021morrigan,
  title={Morrigan: A composite instruction tlb prefetcher},
  author={Vavouliotis, Georgios and Alvarez, Lluc and Grot, Boris and Jim{\'e}nez, Daniel and Casas, Marc},
  booktitle={MICRO-54: 54th Annual IEEE/ACM International Symposium on Microarchitecture},
  pages={1138--1153},
  year={2021}
}

@inproceedings{soundararajan2021pdede,
  title={Pdede: Partitioned, deduplicated, delta branch target buffer},
  author={Soundararajan, Niranjan K and Braun, Peter and Khan, Tanvir Ahmed and Kasikci, Baris and Litz, Heiner and Subramoney, Sreenivas},
  booktitle={MICRO-54: 54th Annual IEEE/ACM International Symposium on Microarchitecture},
  pages={779--791},
  year={2021}
}

@inproceedings{ansari2020divide,
  title={Divide and conquer frontend bottleneck},
  author={Ansari, Ali and Lotfi-Kamran, Pejman and Sarbazi-Azad, Hamid},
  booktitle={2020 ACM/IEEE 47th Annual International Symposium on Computer Architecture (ISCA)},
  pages={65--78},
  year={2020},
  organization={IEEE}
}

@inproceedings{ros2021cost,
  title={A Cost-Effective Entangling Prefetcher for Instructions},
  author={Ros, Alberto and Jimborean, Alexandra},
  booktitle={2021 ACM/IEEE 48th Annual International Symposium on Computer Architecture (ISCA)},
  pages={99--111},
  year={2021},
  organization={IEEE}
}

@inproceedings{cher2001skipper,
  title={Skipper: a microarchitecture for exploiting control-flow independence},
  author={Cher, Chen-Yong and Vijaykumar, TN},
  booktitle={Proceedings. 34th ACM/IEEE International Symposium on Microarchitecture. MICRO-34},
  pages={4--15},
  year={2001},
  organization={IEEE}
}

\end{document}